\PassOptionsToPackage{dvipsnames}{xcolor} % xcolor is used in aastex63; add this here to fix package option-clash errors
\documentclass[twocolumn]{aastex631} %, linenumbers

\usepackage{graphicx}	          % Including figure files
\usepackage{amsmath}	          % Advanced maths commands
\usepackage{upgreek}              % uptau
\usepackage{amssymb}	          % Extra maths symbols
\usepackage{bm}		              % Bold maths symbols, including upright Greek
\usepackage{CJKutf8}              % Chinese name
\usepackage[T1]{fontenc}          % for special char that needs \k command (e.g., Drazkowska)
\usepackage{comment}              % Comment big block
\usepackage{transparent}          % transparent text on top of figure for citation
\usepackage{tabularx}             % Newer table package
\usepackage{threeparttable}       % Table notes smart width
\usepackage{booktabs}             % Better \hline
\usepackage{soul}                 % better underline
\usepackage{ulem}                 % better strike-through
\usepackage{multirow}             % to enable multi-row cells in tables
\usepackage[figuresleft]{rotating}% to rotate figures
\emergencystretch=\maxdimen       % to ease underfull/overfull lines.
\newcommand{\xdex}[1]{\times10^{#1}}

\def\num#1{\numx#1}\def\numx#1e#2{{#1}\mathrm{e}{#2}}

\newcommand{\fracp}[2]{\frac{\partial{#1}}{\partial{#2}}}

\defcitealias{Youdin2005}{YG05}
\defcitealias{Youdin2011}{Y11}
\defcitealias{Takahashi2014}{TI14}
\defcitealias{Takahashi2016}{TI16}
\defcitealias{BL99}{BL99}
\defcitealias{Yang2017}{Y17}
\defcitealias{Carrera2015}{C15}
\defcitealias{Li2018}{L18}

\received{--}
\revised{--}
\accepted{--}
\submitjournal{ApJ}

\shorttitle{Thresholds for Particle Clumping by the SI}
\shortauthors{Li \& Youdin}
\begin{document}
\title{Thresholds for Particle Clumping by the Streaming Instability}
\correspondingauthor{Rixin Li}
\email{rixin.li@cornell.edu}

\author[0000-0001-9222-4367]{Rixin Li
\begin{CJK*}{UTF8}{gbsn}
  (李日新)
\end{CJK*}}
\affiliation{Center for Astrophysics and Planetary Science, Department of Astronomy, 
Cornell University, Ithaca, NY 14853, USA}
\affiliation{Steward Observatory \& Department of Astronomy, University of Arizona, 933 N Cherry Ave, Tucson, AZ 85721, USA}

\author[0000-0002-3644-8726]{Andrew N. Youdin}
\affiliation{Steward Observatory \& Department of Astronomy, University of Arizona, 933 N Cherry Ave, Tucson, AZ 85721, USA}

%%%%%%%%%%%%%%%%%%%%%%%%%%%%%%%%%%%%%%%%%%%%%%%%%%%%%%%%%%%%%%%%%%%%%%%%%%%%%%%%
\begin{abstract}

The streaming instability (SI) is a mechanism to aerodynamically concentrate solids in protoplanetary disks and trigger the formation of planetesimals.  The SI produces strong particle clumping if the ratio of solid to gas surface density -- an effective metallicity -- exceeds a critical value.  This critical value depends on particle sizes and disk conditions such as radial drift-inducing pressure gradients and levels of turbulence.  To quantify these thresholds, we perform a suite of vertically-stratified SI simulations over a range of dust sizes and metallicities.  We find a critical metallicity as low as 0.4\% for the optimum particle sizes and standard radial pressure gradients (normalized value of $\Pi = 0.05$).  This sub-Solar metallicity is lower than previous results due to improved numerical methods and computational effort.  We discover a sharp increase in the critical metallicity for small solids, when the dimensionless stopping time (Stokes number) is $\leq 0.01$.  We provide simple fits to the size-dependent SI clumping threshold, including generalizations to different disk models and levels of turbulence.  We also find that linear, unstratified SI growth rates are a surprisingly poor predictor of particle clumping in non-linear, stratified simulations, especially when the finite resolution of simulations is considered.  Our results widen the parameter space for the SI to trigger planetesimal formation.
\end{abstract}

\keywords{Planet formation (1241); Protoplanetary disks (1300); Planetesimals (1259); Hydrodynamics (1963); Hydrodynamical simulations (767); Gas-to-dust ratio (638)}

%%%%%%%%%%%%%%%%%%%%%%%%%%%%%%%%%%%%%%%%%%%%%%%%%%%%%%%%%%%%%%%%%%%%%%%%%%%%%%%%
\section{Introduction}
\label{sec:intro}

Growth from micron-sized dust to super-kilometer-sized planetesimals constitutes the crucial first step in planet formation \citep{Chiang2010, Johansen2014}.  Our understanding of Solar System bodies, exoplanetary systems, and circumstellar disks is tied to the origin of planetesimals.

Recent observations of protoplanetary disks suggest that the formation of planetesimals and planets is well underway at early evolutionary stages \citep[][]{Andrews2020}.  Diverse substructures, especially dust rings and gaps, are ubiquitous in nearby disks  \citep[][]{ALMAPartnership2015, Andrews2016, Andrews2018, Long2018}.  These rings may be best explained by dust trapping in pressure bumps \citep{Dullemond2018}, which could facilitate planetesimal formation \citep{Pinilla2017, Stammler2019, Carrera2021}.  However, pressure bumps might be caused by already-formed planets \citep{Dong2015, Jin2016, Zhang2018} and kinematic patterns in disk gas have also been attributed to planets \citep{Pinte2018, Teague2019}.  Moreover,  ring substructures have been found in younger Class 0 \& I sources, with ages $\lesssim 0.5$ Myr \citep{Segura-Cox2020, Sheehan2020, Cieza2021}.  If planets are the main cause of observed disk structures, then the first generation planetesimals still need to form in smoother, less structured disks.

The streaming instability (SI) operates in smooth disks, and is one of the leading mechanisms to concentrate solids and rapidly produce planetesimals \citep[hereafter \citetalias{Youdin2005}]{Youdin2005}.  The SI arises from the relative drift of solids and gas in the midplane of protoplanetary disks, and is currently the best studied example of a broader class of astrophysical drag instabilities \citep{Goodman2000, Lin2017, Squire2018}.  High-resolution SI simulations with dust self-gravity produce planetesimals with a broad, top-heavy mass distribution \citep[e.g.,][]{Johansen2015, Simon2016, Schafer2017, Li2019}.  The formation of planetesimals by SI, or related gravitational collapse mechanisms, is supported by Solar System observations, including various properties of comets \citep{Blum2017} and of pristine Kuiper Belt binaries \citep{Nesvorny2010, Nesvorny2019, Nesvorny2021, McKinnon2020}.

The ability of the SI to concentrate particles depends primarily on their size and abundance relative to gas, while the length-scale of particle concentration is set by disk pressure gradients (\citetalias{Youdin2005}; \citealp{Johansen2007}).  Optimal dust sizes for SI concentration have aerodynamic stopping times close to the orbital time, around 1-10 cm in standard disks.  Thus, grain growth by coagulation \citep{Blum2018} is a prerequisite for the SI mechanism.  The midplane (volume) density ratio of dust to gas, $\epsilon$, is the most relevant measure of dust abundance.  However, in real disks -- and simulations with vertical stratification -- $\epsilon$ is controlled by (a) the dust-to-gas surface density ratio, referred to as an effective ``metallicity'' $Z$, and (b) the stirring provided by the SI and other sources of disk turbulence, which balances particle sedimentation.  The amount of stirring by SI in turn scales with disk pressure gradients \citep{Takeuchi2012, Sekiya2018}.

For optimally-sized dust and standard pressure gradients, \citet{Johansen2009a} found that SI triggers strong clumping for $Z \geq 2\%$.  This value is close to Solar abundances \citep{2009Asplund}, implying that with only a modest redistribution of disk solids or removal of disk gas \citep{Youdin2002} the SI might trigger planetesimal formation.  However,  higher values of $Z$ could be needed if accounting for a distribution of particle sizes \citep{Krapp2019, Paardekooper2020, Zhu2021} or additional sources of turbulence \citep{Gole2020}.  

\citet[][hereafter \citetalias{Carrera2015}]{Carrera2015} investigated the size dependence of the SI clumping boundary, and found that  higher $Z$ values are required for particles smaller or larger than the optimum size.  However, the non-linear saturation of the SI, including the details of the clumping boundary, is sensitive to the details of numerical simulations, including domain size, resolution, and boundary conditions \citep{Yang2014, Li2018}.  With numerical improvements, the clumping threshold for very small solids was reduced to lower $Z$ by \citet[][hereafter \citetalias{Yang2017}]{Yang2017}.

Motivated by this previous work,  we revisit the SI clumping boundary across a broad range of particle sizes, using improved numerics and a high level of computational effort for a large suite of simulations.  Our updated results favor the formation of planetesimals via SI clumping at lower metallicities than previously shown.  Our results are especially useful for  planet formation models that include thresholds for planetesimal formation via SI \citep{Drazkowska2014, Schoonenberg2017, Liu2019}.  

The paper is organized as follows.  In Section \ref{sec:method}, we describe our numerical models and simulation setups.  Section \ref{sec:results} details our results, starting with the revised clumping boundary in Section \ref{subsec:boundary}.  Section \ref{subsec:fit} provides fits to the clumping boundary that include generalizations to different disk models and levels of disk turbulence.   Section \ref{subsec:comp_s} compares our clumping results to linear SI growth rates.  In Section \ref{sec:discussion}, we discuss the numerical robustness of this work and compare it to previous works.  Section \ref{sec:conclusions} lays out our conclusions and their implications.

%%%%%%%%%%%%%%%%%%%%%%%%%%%%%%%%%%%%%%%%%%%%%%%%%%%%%%%%%%%%%%%%%%%%%%%%%%%%%%%%
\section{Methods}
\label{sec:method}

To simulate the coupled dynamics of gas and solids in a protoplanetary disk, we use similar methods to \citetalias{Li2018}, which we briefly summarize.  We use the \texttt{ATHENA} code \citep{Stone2008} to model isothermal, un-magnetized gas with the particle module developed by \citet{Bai2010}. 

We use the local shearing box approximation to simulate a vertically stratified small patch of the protoplanetary disk \citep{Hawley1995}.  This patch is centered on a fiducial disk radius ($r$) in the midplane, rotating with the corresponding Keplerian frequency ($\Omega_0$), and described by local Cartesian coordinates $(x, y, z)$ in the radial, azimuthal and vertical directions (respectively).  We apply the standard shearing-periodic boundary conditions (BCs)  in $x$ and $y$.  In the vertical direction, outflow BCs are imposed, where outflowing gas is replenished by density renormalization \citepalias{Li2018}.

The code units of our simulations are $\rho_{\rm g0}$, the midplane gas density; $\Omega_0^{-1}$, the Keplerian orbital timescale, $H$, the vertical gas scale height, and $c_{\rm s} = H \Omega_0$, the sound speed of the gas, which can be applied to an arbitrary disk and location.

Three dimensionless parameters control the physics of the simulations, with values listed in Table \ref{tab:paras_2D}.   The particle stopping time, $t_{\rm s}$ is normalized as
\begin{equation}
    \uptau_{\rm s} = \Omega_0 t_{\rm s}\, ,
\end{equation} 
which measures the strength of coupling between gas and solids \citep{Youdin2010}.  We consider only a single particle size in each simulation, with $\uptau_{\rm s} = 10^{-3}$ to $\uptau_{\rm s} = 1.0$ corresponding to particles sizes of mm to m, respectively, at a few AU in standard disk models \citep{Chiang2010}.  The ratio of the average surface density of solids ($\Sigma_{\rm p}$) to that of gas ($\Sigma_{\rm g} = \sqrt{2 \pi} \rho_{\rm g0} H$) gives our effective metallicity:
\begin{equation}
  Z = \frac{\Sigma_{\rm p}}{\Sigma_{\rm g}}.
\end{equation}
A global radial pressure gradient is parameterized by
\begin{equation}
  \Pi \equiv \frac{\eta v_{\rm K}}{c_{\rm s}} 
\end{equation}
where the applied forcing produces gas rotation that is sub-Keplerian by a fraction $\eta$ of the Keplerian velocity $v_{\rm K} = \Omega_0 r$ (in the absence of particle feedback, \citealp{Nakagawa1986}, \citetalias{Youdin2005}).  

The two most important parameters for vertically stratified SI are $\uptau_{\rm s}$ and the combination $Z/\Pi$ \citep{Sekiya2018} because, when radial pressure gradients force the dynamics, midplane dust-to-gas ratios scale with $Z/\Pi$.   Varying $\Pi$ at fixed $Z/\Pi$ alters the weak effects of gas commpressibility \citep{YJ2007} but otherwise gives similar SI dynamics in terms of the SI length scale of $\eta r$ \citepalias{Youdin2005}.  Thus, we fix $\Pi = 0.05$ without a significant loss of generality.

We initialize the gas density in vertical hydrostatic balance, as a Gaussian of vertical thickness $H$.  We also initialize the particle density about the midplane as a Gaussian of scale height $H_p = \eta r/2$, which gives a midplane particle density,
\begin{align}
  \rho_{\rm p0} &= \frac{\Sigma_{\rm p}}{\sqrt{2\pi} H_{\rm p}}, \label{eq:rho_p0}
\end{align}
of $\rho_{\rm p0} / \rho_{\rm g0} = 2Z/\Pi$ initially.  The initial velocities of gas and particles satisfy the \citet{Nakagawa1986} equilibrium.

We neglect self-gravity for simplicity, as it is not required for strong clumping by the SI.  However, we are motivated by the formation of planetesimals via gravitational collapse \citep{Youdin2002, Johansen2007a}.  Thus, we define particle clumping as strong when the peak particle density $\rho_{\rm p,max}$ exceeds the Roche density, $\rho_{\rm R} = {9\Omega_0^2}/({4\pi G})$.

For a specific value of $\rho_{\rm R}$, we assume a low mass disk with   $Q \equiv c_{\rm s} \Omega_0/(\pi G \Sigma_{\rm g}) =  32$ which gives Roche density 
\begin{equation}
  \rho_{\rm R32} \simeq 180 \rho_{\rm g0}, \label{eq:rho_R32}
\end{equation}
as our adopted strong clumping criterion.  For more massive disks, with lower $Q$ values, planetesimal formation could be triggered with even weaker clumping (and vice-versa for less massive disks).  Fortunately, for cases with strong SI clumping,  $\rho_{\rm p, max}$ tends to greatly exceed our adopted threshold (see Table \ref{tab:prop_2D}).  Thus the precise clumping criterion is unlikely to be significant in most cases.

As shown in Table \ref{tab:paras_2D}, we evolve our 2D $(x-z)$ models for at least $3150\Omega_0^{-1}$ ($\simeq 500$ orbital periods) except for runs with larger particles ($\uptau_{\rm s} \geqslant 0.3$) that have shorter global drift timescales given by \citet[][; see also Column 7 in Table \ref{tab:paras_2D}]{Adachi1976}
\begin{align}
  t_{\rm drift} &= \left(\frac{1+\uptau_{\rm s}^2}{2\Pi\uptau_{\rm s}} \right) \left(\frac{H}{r}\right)^{-1} \Omega_0^{-1} \\
  &= 400 \left(\frac{1+\uptau_{\rm s}^2}{2\uptau_{\rm s}} \right)  \left(\frac{\Pi}{0.05}\right)^{-1} \left(\frac{H/r}{0.05}\right)^{-1} \Omega_0^{-1}\label{eq:t_drift}
\end{align}
The $(\uptau_{\rm s}, Z) = (0.001, 0.04)$ simulation has a slightly shorter simulation time due to numerical stiffness when $\uptau_{\rm s} \ll 1$ and the local dust-to-gas density ratio $\epsilon = \rho_{\rm p}/\rho_{\rm g} \gg 1$ \citep{Bai2010}.  To mitigate the stiffness, we switch to the fully-implicit particle integrator in $\uptau_{\rm s}=0.001$ runs and apply additional measures (see Section \ref{subsubsec:stiffness}) to improve accuracy in this challenging regime.

Our main simulations (all but the double and quadruple resolution tests) have the same high resolution of $\Delta x = \eta r/56 =  H/1280$ (i.e., 1280 grid cells per gas scale height).  
Most simulations have a wide radial domain of $0.8H$
to capture multiple particle filaments and their interactions in a local box.  Only the $\uptau_{\rm s} = 10^{-3}$ simulations have a radially narrower box of $0.2H$ due to computational cost, which was still wide enough to produce multiple clumps.  The vertical extent of our simulation boxes is $0.2H$ for larger particles and $0.4H$ for smaller particles (to account for the possibility of greater stirring of small particles, though this was not in fact seen).  

The number of particles in our simulations is set by
\begin{equation}\label{eq:N_par}
  N_{\rm par} = n_{\rm p} N_{\rm cell} = 2^{20}\times\frac{L_X L_Z}{(0.8\times0.2)H^2},
\end{equation}
where $n_{\rm p} = 4$ is the average number of particles per cell, $N_{\rm cell}$ is the total number of cells, $L_X$ and $L_Z$ are the radial and vertical extent of the box, respectively.  \citet{Bai2010} find that $n_{\rm p} = 1$ gives numerical convergence of unstratified SI simulations, and that grid resolution is more important for non-linear convergence of the particle clumping \citepalias[see also Appendix A in][]{Yang2017}.

For a better measure of particle resolution in stratified simulations, the number of particles should be measured relative to $H_{\rm p}$ not $L_z$, because adding dust-free grid cells away from the midplane does not affect particle resolution.  A difficulty with such a measure is that the precise $H_{\rm p}$ value is a simulation result, not known in advance.  However, we can describe the effective particle resolution more robustly as the number of particles per cell within $\eta r$ of the midplane, i.e.\ $-0.5 < z/(\eta r) < 0.5$, since $H_{\rm p}$ scales with $\eta r$ (absent stirring by additional turbulence).  In this sense, the effective particle resolution in our simulations is 16 (or 32) particles per cell \emph{near the particle midplane layer} in simulations with $L_Z = 0.2H$ (or $0.4 H$ respectively).  This particle resolution should be more than sufficient.

%%%%%%%%%%%%%%%%%%%%%%%%%%%%%%%%%%%%%%%%%%%%%%%%%%%%%%%%%%%%
\begingroup % to localize the effect of the length modification below
\setlength{\medmuskip}{0mu} % to reduce the spacing around binary operators (\times)
\begin{deluxetable*}{lcccccccc}[ht]
  \tablecaption{Simulation Parameters for 2D ($x$-$z$) Models}\label{tab:paras_2D}
  \tablecolumns{9}
  \tablehead{
    \colhead{Run} &
    \colhead{$\uptau_{\rm s}$} &
    \colhead{$Z$} &
    \colhead{$L_X\times L_Z$} &
    \colhead{Resolution} &
    \colhead{$N_{\rm par}$} &
    \colhead{$t_{\rm drift}$} &
    \colhead{$t_{\rm sim}$} &
    \colhead{$\rho_{\rm p,max}>\rho_{\rm R32}$?} \\
    \colhead{} & \colhead{} & \colhead{} &
    \colhead{$H^2$} &
    \colhead{$H^{-1}$} &
    \colhead{} &
    \colhead{$\Omega^{-1}$} &
    \colhead{$\Omega^{-1}$} &
    \colhead{} \\
    \colhead{(1)} & \colhead{(2)} & \colhead{(3)} &
    \colhead{(4)} & \colhead{(5)} & \colhead{(6)} &
    \colhead{(7)} & \colhead{(8)} & \colhead{(9)}
  }
  \startdata
  \hline\hline
  Z1t100    &$1.0$   &$0.01$   &$0.8\times 0.2$ &$1280$ &$2^{20}$ &$400$      & $420$ &Y \\
  Z0.75t100 &$1.0$   &$0.0075$ &$0.8\times 0.2$ &$1280$ &$2^{20}$ &$400$      & $420$ &Y \\
  Z0.6t100  &$1.0$   &$0.006$  &$0.8\times 0.2$ &$1280$ &$2^{20}$ &$400$      & $420$ &Y \\
  Z0.5t100  &$1.0$   &$0.005$  &$0.8\times 0.2$ &$1280$ &$2^{20}$ &$400$      & $420$ &Y \\
  Z0.4t100  &$1.0$   &$0.004$  &$0.8\times 0.2$ &$1280$ &$2^{20}$ &$400$      & $420$ &N \\
  Z1t30     &$0.3$   &$0.01$   &$0.8\times 0.2$ &$1280$ &$2^{20}$ &$727$      & $750$ &Y \\
  Z0.75t30  &$0.3$   &$0.0075$ &$0.8\times 0.2$ &$1280$ &$2^{20}$ &$727$      & $750$ &Y \\
  Z0.6t30   &$0.3$   &$0.006$  &$0.8\times 0.2$ &$1280$ &$2^{20}$ &$727$      & $750$ &Y \\
  Z0.5t30   &$0.3$   &$0.005$  &$0.8\times 0.2$ &$1280$ &$2^{20}$ &$727$      & $750$ &Y \\
  Z0.4t30   &$0.3$   &$0.004$  &$0.8\times 0.2$ &$1280$ &$2^{20}$ &$727$      & $750$ &Y \\
  Z0.3t30   &$0.3$   &$0.003$  &$0.8\times 0.2$ &$1280$ &$2^{20}$ &$727$      & $750$ &N \\
  Z1t10     &$0.1$   &$0.01$   &$0.8\times 0.2$ &$1280$ &$2^{20}$ &$2020$     &$3150$ &Y \\
  Z0.75t10  &$0.1$   &$0.0075$ &$0.8\times 0.2$ &$1280$ &$2^{20}$ &$2020$     &$3150$ &Y \\
  Z0.6t10   &$0.1$   &$0.006$  &$0.8\times 0.2$ &$1280$ &$2^{20}$ &$2020$     &$3150$ &Y \\
  Z0.5t10   &$0.1$   &$0.005$  &$0.8\times 0.2$ &$1280$ &$2^{20}$ &$2020$     &$3150$ &N \\
  Z1t5      &$0.05$  &$0.01$   &$0.8\times 0.4$ &$1280$ &$2^{21}$ &$4010$     &$3150$ &Y \\
  Z0.75t5   &$0.05$  &$0.0075$ &$0.8\times 0.4$ &$1280$ &$2^{21}$ &$4010$     &$3150$ &Y \\
  Z0.6t5    &$0.05$  &$0.006$  &$0.8\times 0.4$ &$1280$ &$2^{21}$ &$4010$     &$3150$ &Y \\
  Z0.5t5    &$0.05$  &$0.005$  &$0.8\times 0.4$ &$1280$ &$2^{21}$ &$4010$     &$3150$ &N \\
  Z1t3      &$0.03$  &$0.01$   &$0.8\times 0.4$ &$1280$ &$2^{21}$ &$6673$     &$3150$ &Y \\
  Z0.75t3   &$0.03$  &$0.0075$ &$0.8\times 0.4$ &$1280$ &$2^{21}$ &$6673$     &$3150$ &Y \\
  Z0.6t3    &$0.03$  &$0.006$  &$0.8\times 0.4$ &$1280$ &$2^{21}$ &$6673$     &$3150$ &Y \\
  Z0.5t3    &$0.03$  &$0.005$  &$0.8\times 0.4$ &$1280$ &$2^{21}$ &$6673$     &$3150$ &N \\
  Z1t2      &$0.02$  &$0.01$   &$0.8\times 0.4$ &$1280$ &$2^{21}$ &$10004$    &$3150$ &Y \\
  Z0.75t2   &$0.02$  &$0.0075$ &$0.8\times 0.4$ &$1280$ &$2^{21}$ &$10004$    &$3150$ &Y \\
  Z0.6t2    &$0.02$  &$0.006$  &$0.8\times 0.4$ &$1280$ &$2^{21}$ &$10004$    &$3150$ &N \\
  Z2t1      &$0.01$  &$0.02$   &$0.8\times 0.4$ &$1280$ &$2^{21}$ &$20002$    &$3150$ &Y \\
  Z1.33t1   &$0.01$  &$0.0133$ &$0.8\times 0.4$ &$1280$ &$2^{21}$ &$20002$    &$3150$ &N \\
  Z1.33t1-2x&$0.01$  &$0.0133$ &$0.2\times 0.4$ &$2560$ &$2^{21}$ &$20002$    &$3150$ &N \\
  Z1.33t1-4x&$0.01$  &$0.0133$ &$0.2\times 0.2$ &$5120$ &$2^{22}$ &$20002$    &$3150$ &N \\
  Z1t1      &$0.01$  &$0.01$   &$0.8\times 0.4$ &$1280$ &$2^{21}$ &$20002$    &$3150$ &N \\
  Z0.75t1   &$0.01$  &$0.0075$ &$0.8\times 0.4$ &$1280$ &$2^{21}$ &$20002$    &$3150$ &N \\
  Z4t0.1    &$0.001$ &$0.04$   &$0.2\times 0.4$ &$1280$ &$2^{19}$ &$2\xdex{5}$&$2952$ &Y\tablenotemark{$\ast$} \\
  Z3t0.1    &$0.001$ &$0.03$   &$0.2\times 0.4$ &$1280$ &$2^{19}$ &$2\xdex{5}$&$3500$ &Y\tablenotemark{$\dag$} \\
  Z2t0.1    &$0.001$ &$0.02$   &$0.2\times 0.4$ &$1280$ &$2^{19}$ &$2\xdex{5}$&$4200$ &N \\
  \enddata
  \tablecomments{Columns: (1) run name (the numbers after Z and t are in units of one hundredth); (2) dimensionless stopping time; (3) metallicity; (4) dimensions of the simulation domain; (5) simulation resolution in units of the number of cells per gas scale height; (6) number of particles (for reference, $2^{20}\approx 1.05\xdex{6}$); (7) global drifting timescale for a test particle (see Equation \ref{eq:t_drift}); (8) total simulation time (9) whether or not the maximum particle density has reached the critical Roche density $\rho_{\rm R32}$.  For all runs: the radial pressure gradient term is $\Pi=0.05$.}
  \vspace{-0.5mm}\tablenotetext{\ast}{This run lasts slightly shorter than the usual $3150/\Omega_0$ because the stiffness issue makes the simulation extremely computational expensive when the particle density becomes high enough (see also Section \ref{subsubsec:stiffness}).}
  \vspace{-0.5mm}\tablenotetext{$\dag$}{In this run, the particle density reaches $\rho_{\rm R32}$ at $t=3161/\Omega_0$, slightly longer than the standard simulation time limit.  We run it a bit longer because the maximum particle density appears to be in a long-term rising trend and is very close to $\rho_{\rm R32}$ at $t=3150/\Omega_0$.}
\end{deluxetable*}
\endgroup

\begin{deluxetable*}{lrrlcrccc}[ht]
  \tablecaption{Time-averaged Disk Properties for 2D $(x-z)$ Models}\label{tab:prop_2D}
  \tablecolumns{9}
  \tablehead{
    \colhead{Run} &
    \colhead{$\rho_{\rm p,max}$} &
    \colhead{$\rho_{\rm p0}$} &
    \colhead{$H_{\rm p}$} &
    \colhead{$t_{\rm sedi-tr}$} &
    \colhead{$t_{\rm pre-cl}$} &
    \colhead{$D_{\mathrm{p},x}$} &
    \colhead{$D_{\mathrm{p},z}$} &
    \colhead{$v_{\mathrm{p},x}$} \\
    \colhead{} &
    \colhead{$\rho_{\rm g0}$} &
    \colhead{$\rho_{\rm g0}$} &
    \colhead{$\eta r$} &
    \colhead{$\Omega^{-1}$} &
    \colhead{$\Omega^{-1}$} &
    \colhead{$c_{\rm s} H$} &
    \colhead{$c_{\rm s} H$} &
    \colhead{$v_{\rm test}$} \\
    \colhead{(1)} & \colhead{(2)} & \colhead{(3)} &
    \colhead{(4)} & \colhead{(5)} & \colhead{(6)} &
    \colhead{(7)} & \colhead{(8)} & \colhead{(9)}
  }
  \startdata
  \hline\hline
  Z1t100    &$ 2892.4$&      - &       - &$35$   &$  22.9$ &               - &               - &     - \\
  Z0.75t100 &$ 1334.8$& $1.02$ &$ 0.147$ &$35$   &$  68.0$ &$\num{4.26e-04}$ &$\num{5.40e-05}$ &$0.49$ \\
  Z0.6t100  &$  854.7$& $0.91$ &$ 0.132$ &$35$   &$ 112.7$ &$\num{4.69e-04}$ &$\num{4.37e-05}$ &$0.51$ \\
  Z0.5t100  &$  422.2$& $0.90$ &$ 0.112$ &$35$   &$  43.0$ &$\num{4.86e-04}$ &$\num{3.12e-05}$ &$0.50$ \\
  Z0.4t100  &$  174.3$& $0.72$ &$ 0.111$ &$35$   &$  76.4$ &$\num{4.11e-04}$ &$\num{3.05e-05}$ &$0.59$ \\
  Z1t30     &$ 3808.8$&      - &       - &$45$   &$  43.3$ &               - &               - &     - \\
  Z0.75t30  &$ 2353.4$& $0.60$ &$  0.25$ &$45$   &$  73.5$ &$\num{2.23e-04}$ &$\num{4.70e-05}$ &$0.68$ \\
  Z0.6t30   &$ 2097.3$& $0.52$ &$ 0.232$ &$45$   &$  82.4$ &$\num{4.50e-04}$ &$\num{4.05e-05}$ &$0.66$ \\
  Z0.5t30   &$ 1402.3$& $0.40$ &$  0.25$ &$45$   &$ 215.1$ &$\num{2.15e-04}$ &$\num{4.67e-05}$ &$0.76$ \\
  Z0.4t30   &$ 1276.9$& $0.35$ &$ 0.226$ &$45$   &$ 355.2$ &$\num{1.72e-04}$ &$\num{3.83e-05}$ &$0.80$ \\
  Z0.3t30   &$   56.6$& $0.30$ &$ 0.203$ &$45$   &       - &$\num{1.51e-04}$ &$\num{3.10e-05}$ &$0.88$ \\
  Z1t10     &$ 3114.2$& $0.68$ &$ 0.293$ &$50$   &$ 279.3$ &$\num{9.07e-05}$ &$\num{2.15e-05}$ &$0.82$ \\
  Z0.75t10  &$  841.3$& $0.51$ &$ 0.294$ &$50$   &$ 327.6$ &$\num{1.10e-04}$ &$\num{2.16e-05}$ &$0.87$ \\
  Z0.6t10   &$  452.3$& $0.40$ &$ 0.297$ &$50$   &$1116.4$ &$\num{1.01e-04}$ &$\num{2.21e-05}$ &$0.97$ \\
  Z0.5t10   &$   27.7$& $0.34$ &$ 0.291$ &$50$   &       - &$\num{9.70e-05}$ &$\num{2.12e-05}$ &$1.02$ \\
  Z1t5      &$ 1987.9$& $0.77$ &$ 0.259$ &$70$   &$ 257.0$ &$\num{4.16e-05}$ &$\num{8.41e-06}$ &$0.72$ \\
  Z0.75t5   &$  861.3$& $0.55$ &$ 0.275$ &$70$   &$ 624.2$ &$\num{4.45e-05}$ &$\num{9.42e-06}$ &$0.83$ \\
  Z0.6t5    &$  186.1$& $0.41$ &$ 0.294$ &$70$   &$1687.8$ &$\num{4.04e-05}$ &$\num{1.08e-05}$ &$0.95$ \\
  Z0.5t5    &$    6.6$& $0.34$ &$ 0.295$ &$70$   &       - &$\num{3.98e-05}$ &$\num{1.09e-05}$ &$1.01$ \\
  Z1t3      &$  428.6$& $1.03$ &$ 0.194$ &$120$  &$ 603.0$ &$\num{2.16e-05}$ &$\num{2.83e-06}$ &$0.53$ \\
  Z0.75t3   &$ 1236.8$& $0.67$ &$ 0.224$ &$120$  &$ 696.3$ &$\num{1.64e-05}$ &$\num{3.78e-06}$ &$0.67$ \\
  Z0.6t3    &$  413.7$& $0.50$ &$ 0.241$ &$120$  &$2363.2$ &$\num{1.47e-05}$ &$\num{4.36e-06}$ &$0.76$ \\
  Z0.5t3    &$    5.0$& $0.39$ &$ 0.255$ &$120$  &       - &$\num{1.18e-05}$ &$\num{4.88e-06}$ &$0.84$ \\
  Z1t2      &$ 2081.3$& $1.18$ &$  0.17$ &$150$  &$2771.5$ &$\num{5.92e-06}$ &$\num{1.44e-06}$ &$0.49$ \\
  Z0.75t2   &$ 1051.0$& $0.79$ &$  0.19$ &$150$  &$1869.7$ &$\num{5.07e-06}$ &$\num{1.80e-06}$ &$0.59$ \\
  Z0.6t2    &$    4.5$& $0.56$ &$ 0.214$ &$150$  &       - &$\num{3.32e-06}$ &$\num{2.30e-06}$ &$0.70$ \\
  Z2t1      &$  812.6$& $3.54$ &$ 0.113$ &$250$  &$ 846.7$ &$\num{1.40e-06}$ &$\num{3.20e-07}$ &$0.31$ \\
  Z1.33t1   &$   10.1$& $2.18$ &$ 0.122$ &$250$  &       - &$\num{1.31e-06}$ &$\num{3.74e-07}$ &$0.40$ \\
  Z1.33t1-2x&$    7.7$& $1.86$ &$ 0.143$ &$250$  &       - &$\num{1.47e-06}$ &$\num{5.12e-07}$ &$0.41$ \\
  Z1.33t1-4x&$    6.4$& $1.27$ &$ 0.209$ &$250$  &       - &$\num{2.36e-06}$ &$\num{1.10e-06}$ &$0.48$ \\
  Z1t1      &$    7.5$& $1.56$ &$ 0.128$ &$250$  &       - &$\num{1.57e-06}$ &$\num{4.12e-07}$ &$0.45$ \\
  Z0.75t1   &$    6.0$& $1.06$ &$ 0.142$ &$250$  &       - &$\num{1.67e-06}$ &$\num{5.03e-07}$ &$0.52$ \\
  Z4t0.1    &$  493.3$& $6.20$ &$ 0.129$ &$2100$ &$2575.1$ &$\num{2.27e-07}$ &$\num{4.17e-08}$ &$0.34$ \\
  Z3t0.1    &$  327.0$& $3.48$ &$ 0.172$ &$2200$ &$2955.0$ &$\num{2.35e-07}$ &$\num{7.43e-08}$ &$0.44$ \\
  Z2t0.1    &$    9.7$& $1.89$ &$ 0.212$ &$2500$ &       - &$\num{2.63e-07}$ &$\num{1.12e-07}$ &$0.55$ \\
  \enddata
  \tablecomments{ Columns: 
  (1) run name (the numbers after Z and t are in units of one hundredth); 
  (2) maximum particle density in simulations; 
  (3) midplane particle density; 
  (4) particle scale height; 
  (5) end time of the transient sedimentation by eye; 
  (6) end time of the pre-clumping phase (i.e., the first time when $\rho_{\rm p,max}$ reaches $(2/3)\rho_{\rm R32}$; 
  (7) radial diffusion coefficient extracted from $0.5 \partial \sigma_x^2 / \partial t$; 
  (8) vertical particle diffusion coefficient derived from $\langle H_{\rm p} \rangle_t^2 \uptau_{\rm s}$; 
  (9) ratio between radial particle velocity and test particle velocity, $-2\eta v_{\rm K} \uptau_{\rm s} / (1 + \uptau_{\rm s}^2)$. }
  \tablecomments{The time average is taken over the saturated pre-clumping phase, that is, after the transient sedimentation (including the initial settling and the subsequent bounce of $H_{\rm p}$) and before the first time that $\rho_{\rm p,max}$ exceeds $(2/3)\rho_{\rm R32}$.}
\end{deluxetable*}

%%%%%%%%%%%%%%%%%%%%%%%%%%%%%%%%%%%%%%%%%%%%%%%%%%%%%%%%%%%%
\begin{figure*}[tbh!]
  \centering
  \includegraphics[width=0.8\linewidth]{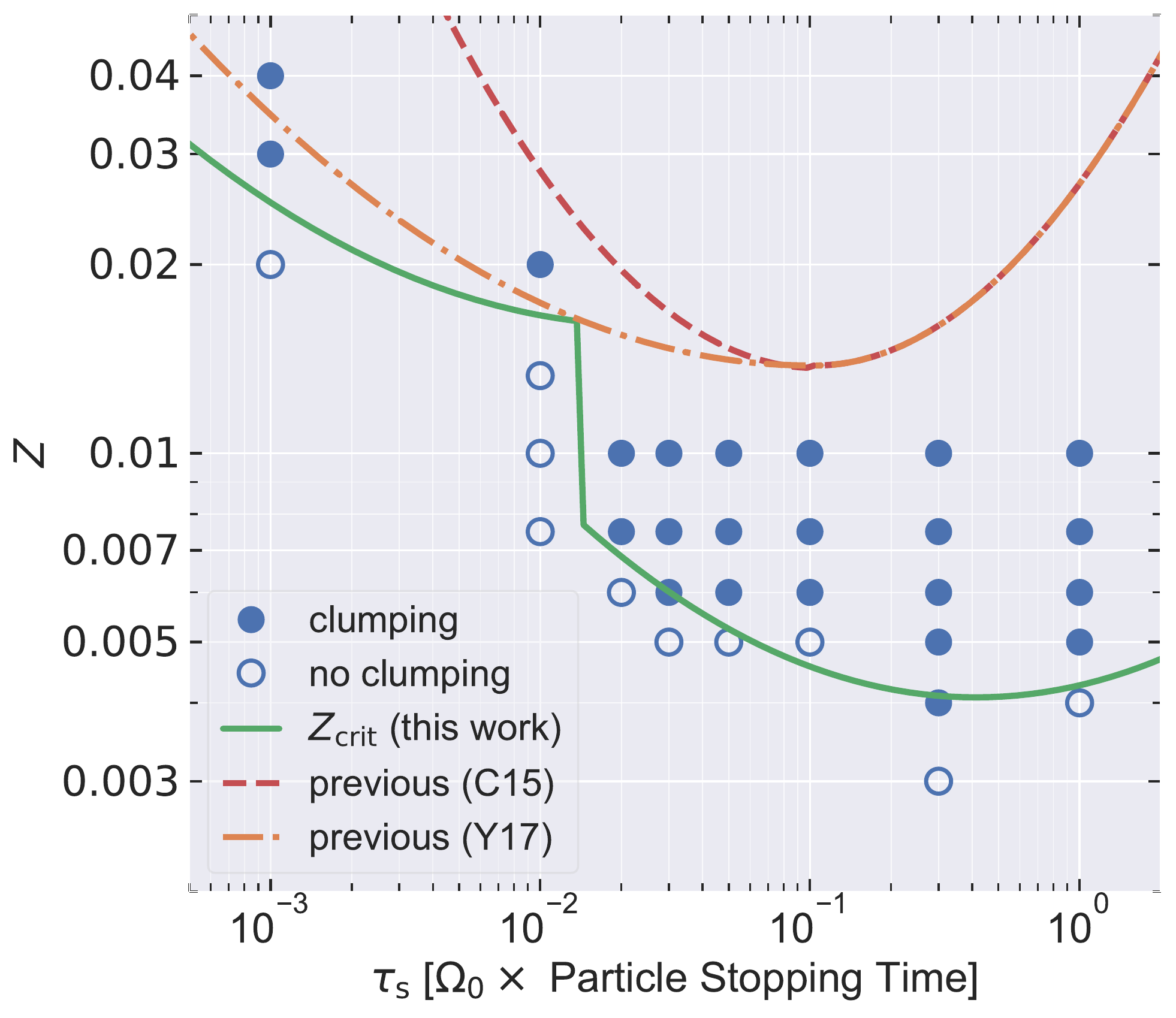}
  \caption{Overview of particle clumping by the SI for runs of different stopping times, $\uptau_{\rm s}$, and metallicities, $Z$, with radial pressure gradients set by $\Pi = 0.05$, and no additional sources of turbulence.  Strong clumping occurs (or does not occur) for cases with filled (or open) circles.  The minimum $Z$ for strong clumping is fit by the \textit{solid curve}.  Previous boundaries are shown by the \textit{dashed} and \textit{dash-dotted} curves.  See text for additional discussion including generalizations of the clumping boundary in Section \ref{subsec:fit}.  Table \ref{tab:paras_2D} gives additional simulation details.  
  An interactive version of this plot with detailed analysis (i.e., visualizations similar to Figures \ref{fig:rhop_xt_321}, \ref{fig:eg_H_p_epsilon}, \ref{fig:rhop_xz_tauZ}, and \ref{fig:eg_s_max}) of each simulation is available at  \href{https://rixinli.com/tauZmap.html}{https://rixinli.com/tauZmap.html} and in the online version of the article. \label{fig:map}}
  %% CODE BELOW adds transparent clickable citations to C15 and Y17 (reveal them with \transparent{0.5})
  \vspace{-5.8cm} \hspace{-0.95cm} %% jump back to the fine-tuned legend position
  \transparent{0.0}{\Large \citetalias{Carrera2015}}
  \hspace{0.95cm} \vspace{5.375cm} \\
  \vspace{-5.12cm} \hspace{-0.975cm}
  \transparent{0.0}{\Large \citetalias{Yang2017}}
  \hspace{0.975cm} \vspace{5.55cm} \\ %% jump forward so normal text won't follow
  \vspace{-\baselineskip}\vspace{-\baselineskip}
\end{figure*}

\begin{figure*}[tbh!]
  \centering
  \includegraphics[width=\linewidth]{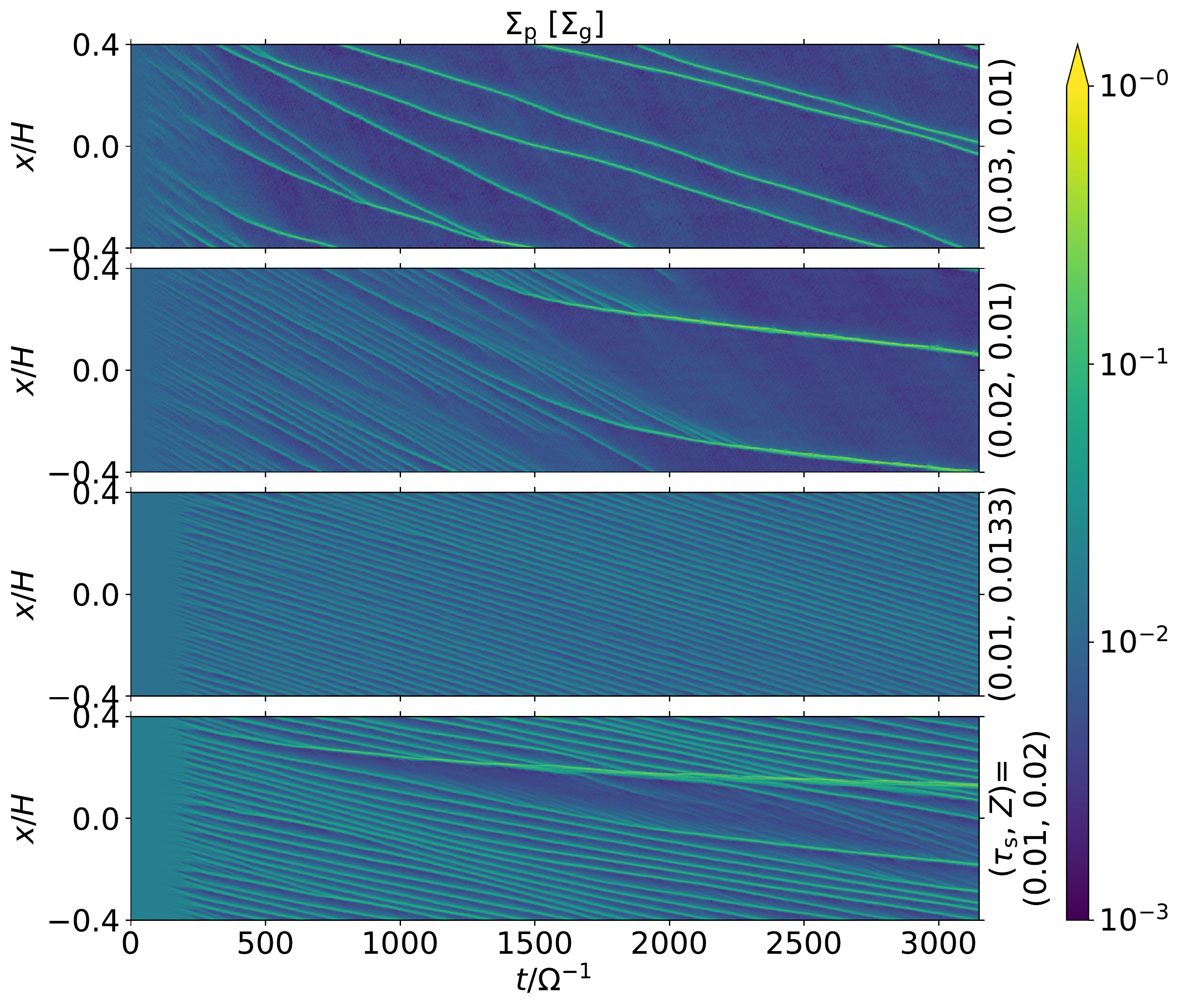}
  \caption{Comparison of the radial concentration of particles as a function of time for Runs \texttt{Z1t3}, \texttt{Z1t2}, \texttt{Z1.33t1}, and \texttt{Z2t1} (\textit{from top to bottom}), showing different ways that strong particle clumping emerges.  For $\uptau_{\rm s} \geq 0.02$, (\emph{top two panels}) strong clumping emerges as transient clumps dissolve and feed a few dominant filaments.  On the contrary, at $\uptau_{\rm s} = 0.01$, initial weak clumping is uniformly spaced and long lived (\emph{bottom two panels}).  Only at higher $Z$ ($ = 0.02$, \emph{fourth panel}) do the filaments merge to produce strong clumping.
  \label{fig:rhop_xt_321}}
\end{figure*}

%%%%%%%%%%%%%%%%%%%%%%%%%%%%%%%%%%%%%%%%%%%%%%%%%%%%%%%%%%%%%%%%%%%%%%%%%%%%%%%%
\section{Simulation Results}
\label{sec:results}

We conduct a suite of axisymmetric, vertically-stratified SI simulations with parameters listed in Table \ref{tab:paras_2D}.  Figure \ref{fig:map} shows which runs produce strong particle clumping and which do not.  The clumping is considered to be strong if the maximum local particle density reaches the Roche density (for a fiducial disk, see Equation \ref{eq:rho_R32} and related discussion).  We define $Z_{\rm crit}$, as the critical metallicity above which strong particle clumping is found in SI simulations, for fixed values of other parameters.  Table \ref{tab:prop_2D} gives summary statistics of the simulations.

In this section, we present our results for $Z_{\rm crit}$ in Section \ref{subsec:boundary} and examine average midplane dust-to-gas density ratios in Section \ref{subsec:rhop}.  Section \ref{subsec:fit} provides a revised fitting formula for $Z_{\rm crit}(\uptau_{\rm s})$.  These fits address both pure SI, a generalization for additional sources of turbulence, and other choices of the pressure support parameter (that differ from the $\Pi = 0.05$ of the simulations).  Section \ref{subsec:diff_turb} measures turbulent diffusion from SI and Section \ref{subsec:comp_s} compares our clumping results to the linear growth rates of unstratified SI.

%%%%%%%%%%%%%%%%%%%%%%%%%%%%%%%%%%%%%%%%%%%%%%%%%%%%%%%%%%%%
\subsection{Boundary for Strong Particle Clumping}
\label{subsec:boundary}

The particle clumping behavior in Figure \ref{fig:map} shows two remarkable features.  First, for intermediate $\uptau_{\rm s} > 0.01$, particle clumping occurs at lower values of $Z$, i.e. $Z_{\rm crit}$ is lower, than in previous work \citepalias{Carrera2015}.\footnote{The \citetalias{Yang2017} curves for $\uptau_{\rm s} > 0.01$ use the  \citetalias{Carrera2015} simulation results.}
The simplest reason for the difference is that \citetalias{Carrera2015} investigate clumping while increasing $Z$ in time, which significantly reduces computational cost.  By performing a larger number of fixed-$Z$ simulations, we allow more time for clumping to develop at lower $Z$.  Section \ref{subsec:comp_pre} compares to previous works on $Z_{\rm crit}$ in more detail.

Particle sizes near $\uptau_{\rm s} = 0.3$ produce clumping for the lowest amount of solids $Z = 0.4\%$.  Many simulations of the SI have adopted $\uptau_{\rm s} \simeq 0.3$ \citep{Johansen2012, Yang2014, Schafer2017, Li2018}, which is clearly a favorable assumption for strong particle clumping and planetesimal formation.

The second remarkable feature in Figure \ref{fig:map} is a sharp decrease of the critical metallicity $Z_{\rm crit}$ when $\uptau_{\rm s}$ increases from $0.01$ to $0.02$.  Variations of  $Z_{\rm crit}$ vs.\ $\uptau_{\rm s}$ are much more gradual elsewhere, i.e.\ for both $\uptau_{\rm s} < 0.01$ and $>0.02$.  

To help understand this transition, Figure \ref{fig:rhop_xt_321} illustrates how particles  clump radially over time, for $\uptau_{\rm s} = 0.01$ vs.\ $0.02$ and $0.03$.  For $\uptau_{\rm s} = 0.01$, clumps form evenly spaced with similar, low amplitudes that persist for many orbits.  Only at sufficiently high $Z$ ($\geq 0.02$) is this uniform arrangement  disrupted, so that mergers of filaments result in strong clumping.  By contrast, for $\uptau_{\rm s} = 0.02$ and $0.03$, clumping is initially less uniform with unequal amplitudes, and most clumps are transient.  This more stochastic clumping readily leads to clump mergers and the emergence of strong clumping at lower $Z$.

The ultimate reason why this transition to more stochastic clumping happens at $\uptau_{\rm s} > 0.01$ remains elusive.  We searched for other quantities that vary significantly across this $Z_{\rm crit}$ transition, but none were identified, as we detail in the subsequent analyses of this section.

Our clumping boundary for $\uptau_{\rm s}  = 0.001$ -- $0.01$ agrees well, if not precisely, with the current state-of-the-art results by \citetalias{Yang2017} (see Section \ref{subsec:comp_pre} for more discussion).  This agreement is encouraging since numerical stiffness is a significant challenge for  $\uptau_{\rm s} \ll 1$ simulations, as explained in Section \ref{subsubsec:stiffness}.  Since \citetalias{Yang2017} did not revisit simulations with $\uptau_{\rm s} > 0.01$, they were not sensitive to the $Z_{\rm crit}$ transition reported here. 

%%%%%%%%%%%%%%%%%%%%%%%%%%%%%%%%%%%%%%%%%%%%%%%%%%%%%%%%%%%%
\begin{figure}[tbh!]
  \centering
  \includegraphics[width=\linewidth]{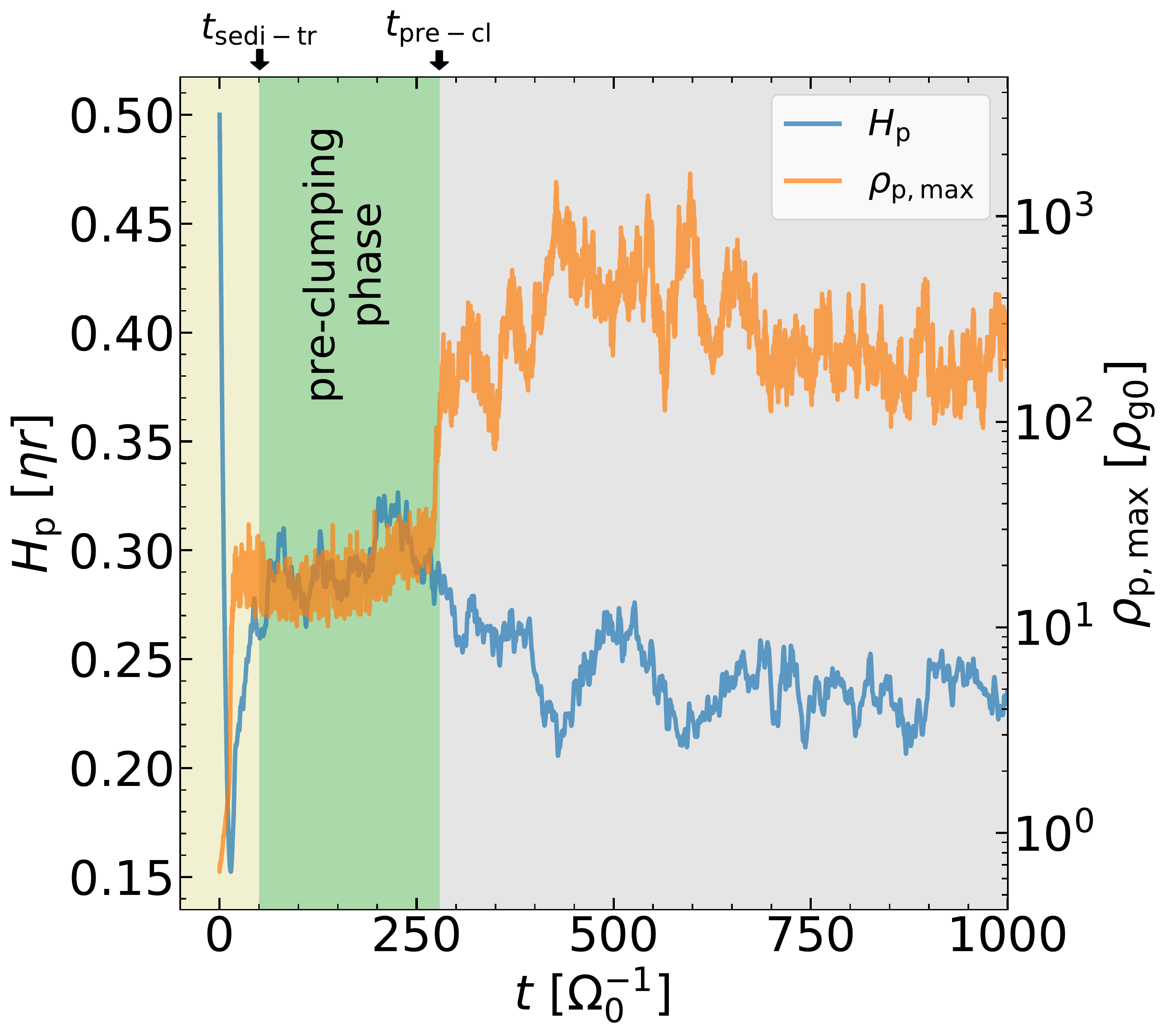}
  \caption{Evolution of the particle scale-height and midplane density from Run \texttt{Z1t10}.  This simulation (and most others) can be divided into three phases: transient sedimentation phase, pre-clumping phase, and strong clumping phase, as discussed in the text.  \label{fig:eg_H_p_epsilon}}
\end{figure}

\begin{figure*}[tbh!]
  \centering
  \includegraphics[width=\linewidth]{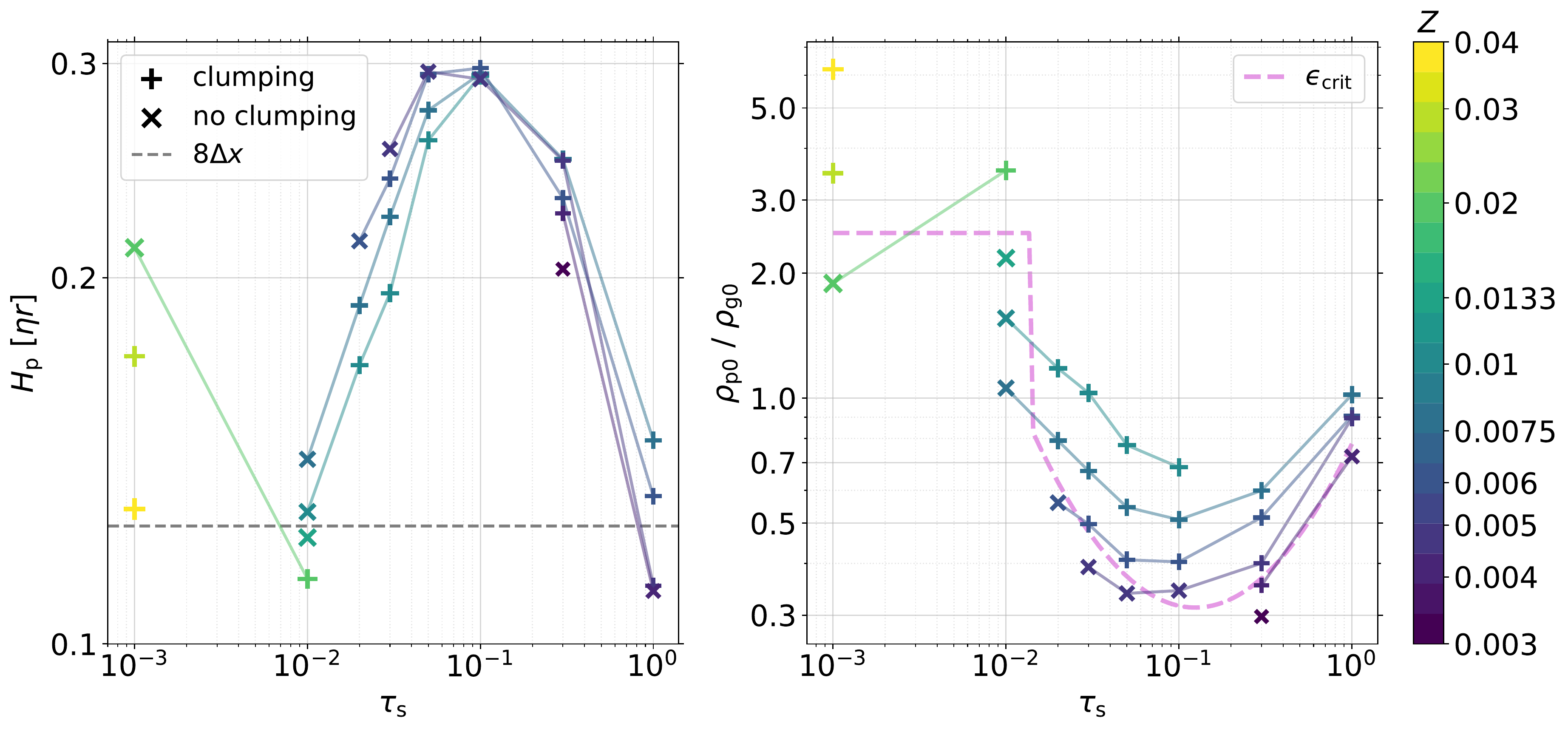}
  \caption{The particle scale height (\textit{left}) and the 
  midplane particle-to-gas density ratio (\textit{right}), time-averaged over the pre-clumping phase, for all runs.  Cases that produce strong particle clumping (\textit{plus}) are differentiated with those that do not (\textit{cross}).  Runs are color-coded by $Z$ values, and simulations with the same metallicity are connected with line segments with the corresponding color.  A black dashed reference line (left panel) shows the length scale $8\Delta x$ for comparison.  A magenta dashed curve (right panel) shows our approximate fit to the clumping threshold $\epsilon_{\rm crit}$.  \label{fig:H_p_epsilon}}
\end{figure*}

\begin{figure*}[tbh!]
  \centering
  \includegraphics[width=\linewidth]{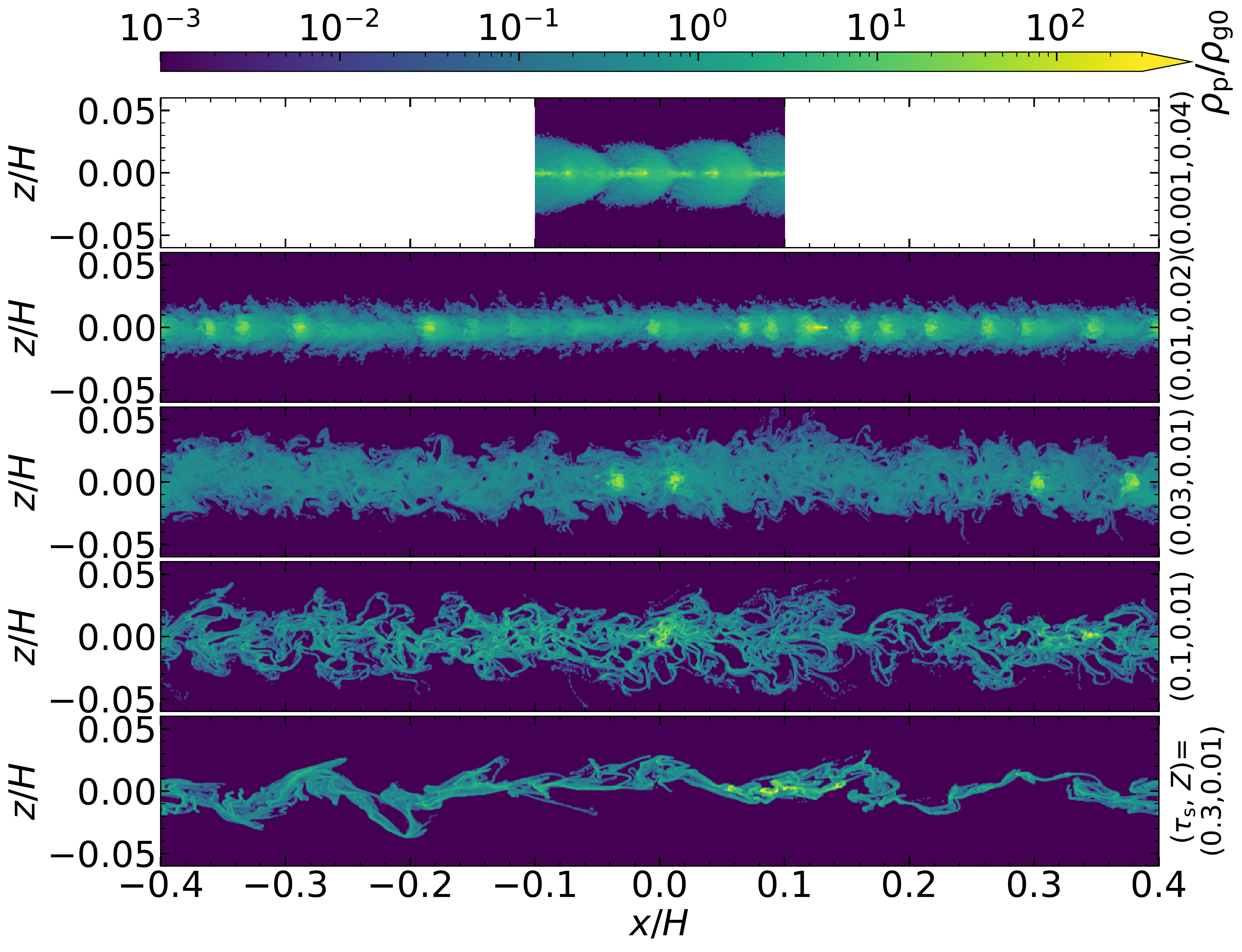}
  \caption{Final snapshots of particle density for a range of particle sizes, increasing from top to bottom in clumping Runs \texttt{Z4t01}, \texttt{Z2t1}, \texttt{Z1t3}, \texttt{Z1t10}, and \texttt{Z1t30}.  
  The diverse structures (discussed in the text) correlate with the irregular variations of $H_{\rm p}$ with $\uptau_{\rm s}$ as shown in Figure \ref{fig:H_p_epsilon}.  Note that the vertical gas domain is larger than the range of these plots. 
  \label{fig:rhop_xz_tauZ}}
\end{figure*}

\begin{figure}[tbh!]
  \centering
  \includegraphics[width=\linewidth]{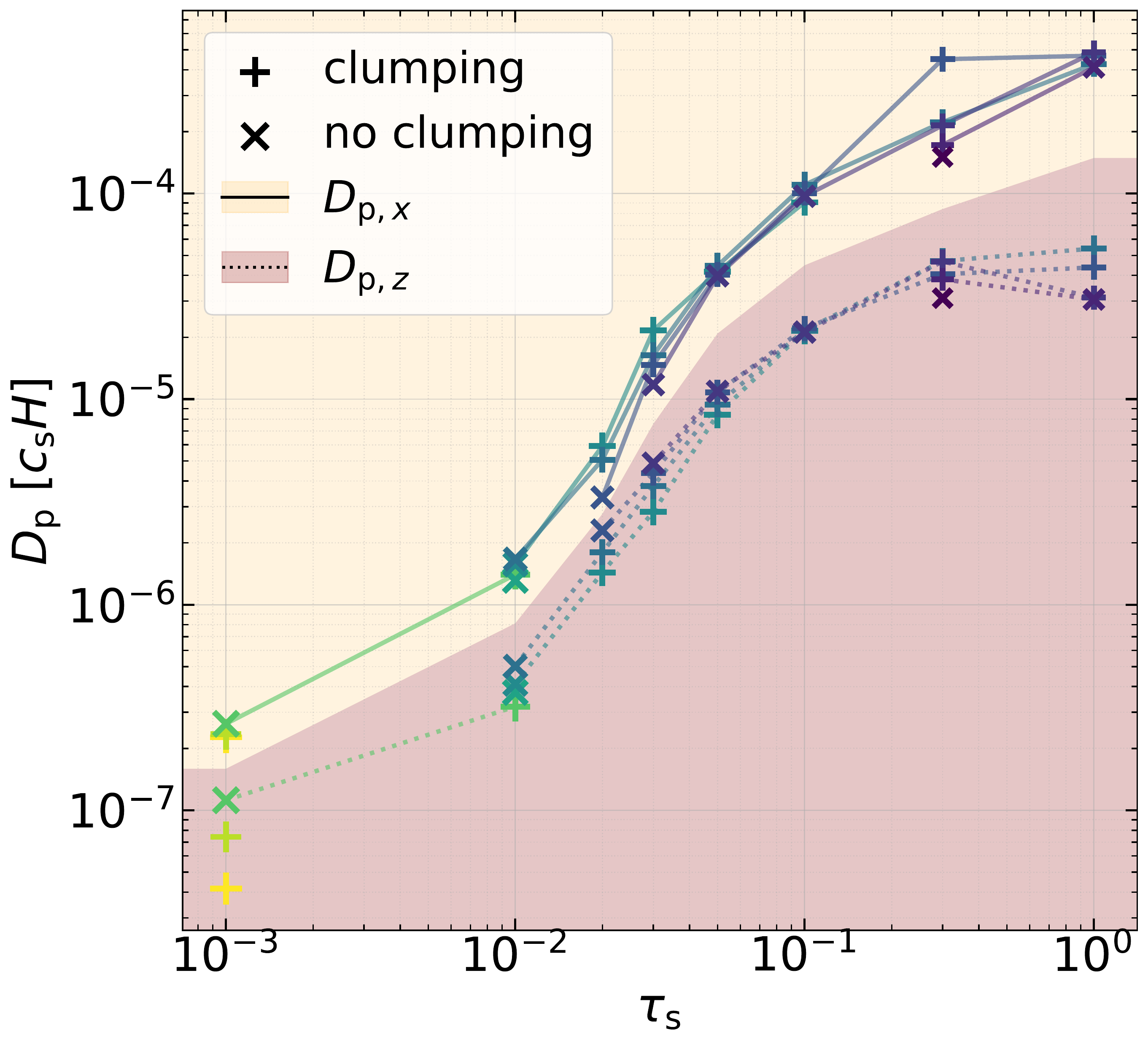}
  \caption{Diffusion coefficients measured in our simulations. Radial coefficients (connected by straight lines against a yellow background) and vertical coefficients (connect by dotted lines against a pink background) follow the same trend with $\uptau_{\rm s}$.  Symbols, lines and their colors have the same meaning as Figure \ref{fig:H_p_epsilon}.  The diffusivities for $\uptau_{\rm s} = 1$ are of order the dimensional estimate $D_{\rm p} = (\eta r)^2 \Omega = \num{2.5e-4}\ c_{\rm s} H$. See text for further discussion. \label{fig:D_p}}
\end{figure}

%%%%%%%%%%%%%%%%%%%%%%%%%%%%%%%%%%%%%%%%%%%%%%%%%%%%%%%%%%%%
\subsection{Particle Scale-Height and Midplane Density}
\label{subsec:rhop}

Particle settling competes with stirring by the gas to establish the particle scale height $H_{\rm p}$ and midplane density $\rho_{\rm p0} = \epsilon \rho_{\rm g0}$.  The density ratio $\epsilon$ describes the strength of drag feedback and is a fundamental parameter of unstratified SI \citepalias{Youdin2005}.  

The condition for particle clumping by the SI is often approximated as $\epsilon \gtrsim 1$ in the midplane, due to the increase in linear SI growth rates with $\epsilon$, which is strong across $\epsilon = 1$ \emph{if} $\uptau_{\rm s} \ll 1$  (\citetalias{Youdin2005}, \citealp{YJ2007}).  This subsection refines the minimum $\epsilon$ needed for SI clumping with stratification.\footnote{Since \citet{Johansen2007} found strong particle clumping for $\epsilon = 0.2$ and $\uptau_{\rm s} = 1.0$, the $\epsilon > 1$ criterion is known to be approximate in unstratified SI simulations.}

To measure $H_{\rm p}$, $\epsilon$ and other quantities, we divide our simulations into three phases, illustrated in Figure \ref{fig:eg_H_p_epsilon}.  First, $t < t_{\rm sedi-tr}$ is the ``transient sedimentation'' phase, where particles settle towards, then bounce away from, the midplane.  Next, $t_{\rm sedi-tr} < t < t_{\rm pre-cl}$ is a quasi-equilibrium ``pre-clumping'' phase, before the ``strong clumping'' phase which begins if and when $\rho_{\rm p, max}$ first exceeds (2/3)$\rho_{\rm R32}$.   Table \ref{tab:prop_2D} lists these transition times ($t_{\rm sedi-tr}$, $t_{\rm pre-cl}$) for all of our simulations. 

Our quantitative analyses mainly consider the pre-clumping phase, as this helps to understand the properties of dust layers -- including $H_{\rm p}$ and $\rho_{\rm p0}$ --  that are capable (or not) of producing strong particle clumping.   Only two cases (\texttt{Z1t100} and \texttt{Z1t30}) produce strong particle concentrations too quickly ($t_{\rm sedi-tr} > t_{\rm pre-cl}$), and lack a pre-clumping phase.  These cases are omitted and are far from the clumping boundary, our main interest. 

We measure $H_{\rm p}$ as the standard deviation (i.e.\ square root of the variance) of vertical particle positions. 
Figure \ref{fig:H_p_epsilon} (left panel) presents $H_{\rm p}$ measured during the pre-clumping phase (see also Table \ref{tab:prop_2D}).  The characteristic $H_{\rm p} \sim 0.2 \eta r$, with variations about that value that have a complicated $\uptau_{\rm s}$ and $Z$ dependence.  We now describe $H_{\rm p}(\uptau_{\rm s}, Z)$ in the vicinity of the clumping boundary.

We first consider the variation of $H_{\rm p}$  with $\uptau_{\rm s}$ at fixed $Z$, i.e.\ $\partial H_{\rm p} / \partial \uptau_{\rm s}$.  A flat $\partial H_{\rm p} / \partial \uptau_{\rm s} = 0$ would arise if the greater settling of larger solids was precisely balanced by the effect of larger solids stirring stronger turbulence.  Instead we have an interval from $0.01 < \uptau_{\rm s} < 0.1$ where $\partial H_{\rm p} / \partial \uptau_{\rm s} > 0$ and greater stirring by larger solids is the stronger effect.  And we have two regions, $\uptau_{\rm s} > 0.1$ and $\uptau_{\rm s} < 0.01$,  where the greater settling of larger solids is the stronger effect.  These $H_{\rm p}$ trends are interpreted in terms of effective particle diffusion rates in Section \ref{subsec:diff_turb} (see Figure \ref{fig:D_p}).  

This complex $H_{\rm p}$ dependence on $\uptau_{\rm s}$ is reflected in different morphologies of the midplane particle layer,  shown in Figure \ref{fig:rhop_xz_tauZ}.  As particle size increases, the vertical structure of the particle layers transitions from ``fish-shaped" structures (for $\uptau_{\rm s} = 0.001$), to a layer that is smoothly ``feathered" at the surface (for $\uptau_{\rm s} = 0.01$), then developing more prominent low-density cavities (for $\uptau_{\rm s} = 0.03$ and $0.1$, similar to the unstratified runs with $\uptau_{\rm s} = 0.1$ in \citealp{Johansen2007}), and finally a corrugated structure with significant asymmetries about the midplane (for $\uptau_{\rm s} = 0.3$, a structure familiar from previous 3D, stratified SI simulations such as \citealp{Johansen2009}; \citetalias{Li2018}).  Especially at lower $\uptau_{\rm s}$, higher resolution 3D simulations are needed to confirm these structures.  While the origins of these  structures are mysterious, their diversity is consistent with complex $H_p(\uptau_{\rm s})$ trends.

The metallicity dependence of $H_{\rm p}$ is slightly simpler.  For $\uptau_{\rm s} > 0.1$, $\partial H_{\rm p} / \partial Z >0$ so that more solids trigger greater stirring.  But for  $\uptau_{\rm s} < 0.1$, $\partial H_{\rm p} / \partial Z <0$, meaning that the increased inertia of solids reduces stirring in this regime.  Note that all of these $H_{\rm p}(\uptau_{\rm s}, Z)$ trends hold near the clumping boundary, where our simulations were performed.

We measure the midplane particle density by assuming a Gaussian vertical distribution with the width of the measured $H_{\rm p}$, using Equation \ref{eq:rho_p0}.
\footnote{Direct measurements of the dust density give very similar results, but use an arbitrary, and usually resolution-dependent, averaging length.  Vertical $\rho_{\rm p}$ profiles (averaged in time and radius) are  confirmed to be nearly Gaussian.}  
Figure \ref{fig:H_p_epsilon} (right panel) shows the midplane density ratios $\epsilon = \rho_{\rm p0}/\rho_{\rm g0}$ during the pre-clumping phase.  For higher $Z$, $\epsilon$ increases, as expected.  At fixed $Z$, $\epsilon \propto 1/H_{\rm p}$ reflects the complex $\uptau_{\rm s}$ dependence that was already described for $H_{\rm p}$.

The critical density ratio for strong clumping, $\epsilon_{\rm crit}$, is the dividing line between clumping and non-clumping (i.e.\ weak clumping) cases in Figure \ref{fig:H_p_epsilon} (right panel).  This $\epsilon_{\rm crit}$ is not always unity and varies by almost an order of magnitude across the $\uptau_{\rm s}$ values considered.  The lowest $\epsilon_{\rm crit} \simeq 0.35$ at $\uptau_{\rm s} = 0.3$ and the highest $\epsilon_{\rm crit} \simeq 2.5$ at $\uptau_{\rm s} = 0.01$ and 0.001.  The jump in $\epsilon_{\rm crit}$ from $\uptau_{\rm s} = 0.02$ to $0.01$ is even larger than the corresponding jump in $Z_{\rm crit}$ (since $H_{\rm p}$ is smaller for $\uptau_{\rm s} = 0.01$).  This result makes the jump even more surprising, from the perspective that $\epsilon$ is more fundamental to the SI. 

These variations in $\epsilon_{\rm crit}$ demonstrate that the assumption of $\epsilon_{\rm crit} > 1$ for SI clumping should be taken with caution, especially for smaller particles where a higher dust-to-gas ratio is required.  Furthermore, our result of $\epsilon_{\rm crit} < 1$ for $0.02 \lesssim \uptau_{\rm s} \lesssim 1.0$ is consistent with \citet{Gole2020}.  With $\uptau_{\rm s} = 0.3$, they found $\epsilon_{\rm crit} \simeq 0.5$ in 3D SI simulations with a forced turbulence characterized by a dimensionless strength $\alpha =  10^{-3.5}$, similar to the $\epsilon_{\rm crit} \simeq 0.35$ found here (see Section \ref{subsubsec:alphathresh} for more discussion of forced turbulence).  We caution again that reported dust-gas ratios apply only to the single size bin considered, not to a broad distribution of particle sizes.

%%%%%%%%%%%%%%%%%%%%%%%%%%%%%%%%%%%%%%%%%%%%%%%%%%%%%%%%%%%%
\subsection{Fits to the Clumping Boundary}
\label{subsec:fit}

Approximate, analytic fits to the SI particle clumping boundary are useful for estimating when and where SI-triggered planetesimal formation could occur in protoplanetary disks.  A well-motivated analytic fit allows the results of simulations with specific parameters to be applied to a wide range of actual disk conditions.  Note that our use of a single particle size cannot account for size distributions, which remains a goal of future work.   

In Section \ref{subsubsec:Zcrit}, we fit the clumping criterion for pure stratified SI, accounting for radial pressure gradients that differ from the value ($\Pi = 0.05$) in our simulations.  In Section \ref{subsubsec:alphathresh}, we include the effect of additional turbulence, described by a dimensionless $\alpha$-parameter. 

Strong clumping by these criteria does not guarantee gravitational collapse into planetesimals, especially in lower mass disks (see Section \ref{sec:method}).  While more simulations of collapse near SI clumping boundaries are needed, most of our strong clumping runs exceeded the clumping threshold by a significant margin in long-lived clumps, which favors collapse in most scenarios.

%%%%%%%%%%%%%%%%%%%%%%%%%%%%%%%%%%%%%%%%%%%%%%%%%%%%%%%%%%%%
\subsubsection{Fits to the ``Pure" SI Clumping Boundary}
\label{subsubsec:Zcrit}

To account for the range of pressure gradients in disks, we recall that the fundamental parameter of stratified SI is $Z/\Pi$ (\citealp{Sekiya2018}, see Section \ref{sec:method}). Thus, our simulation results for $Z_{\rm crit}(\Pi = 0.05)$ [now explicitly labelled with our choice of $\Pi$] are generalized as
\begin{align}
    Z_{\rm crit}(\Pi) =  \frac{\Pi}{0.05} Z_{\rm crit}(\Pi = 0.05)\, .
\end{align}
This generalization does not account for pressure bumps where $\Pi \rightarrow 0$, or a small enough value that $\Pi$ is not constant on scales of interest, usually several $\eta r = \Pi H$.  Pressure bumps require a dedicated type of SI simulation \citep{Onishi2017, Carrera2021}, and can also trigger Rossby wave instabilities \citep{Lovelace1999}.

We fit the clumping boundary with a piecewise quadratic function for $\log{(Z_{\rm crit}/\Pi)}$ versus $\log{\uptau_{\rm s}}$.
The function is discontinuous at $\uptau_{\rm s} = 0.015$ as we do not resolve the sharp transition between $\uptau_{\rm s} = 0.01$ -- $0.02$.  After experimenting with different optimization procedures, we ultimately chose the fit parameters by eye, which appropriate for the sparsity of the $\uptau_{\rm s}$ samples (only two for $\uptau_{\rm s} \leq 0.01$ and the somewhat jagged boundary (not exactly obeying any quadratic dividing line) for $\uptau_{\rm s} > 0.01$.  Our preferred fit is
\begin{equation}\label{eq:Z_crit}
  \log{\left(\frac{Z_{\rm crit}}{\Pi}\right)} = A (\log{\uptau_{\rm s}})^2 + B \log{\uptau_{\rm s}} + C,
\end{equation}
where
\begin{align}
  \begin{cases}
    A=0.1,\ B=0.32,\ C=-0.24 & \text{if } \uptau_{\rm s} < 0.015 \\
    A=0.13,\ B=0.1,\ C=-1.07 & \text{if } \uptau_{\rm s} > 0.015
  \nonumber \end{cases}.
\end{align}

%%%%%%%%%%%%%%%%%%%%%%%%%%%%%%%%%%%%%%%%%%%%%%%%%%%%%%%%%%%%
\subsubsection{SI Clumping Threshold with Extra Turbulence}
\label{subsubsec:alphathresh}

The fit in Equation \ref{eq:Z_crit} does not include additional sources of turbulence beyond the dust gas-interactions that drive the SI.  Additional turbulence can provide additional stirring (increase $H_{\rm p}$) and reduce the midplane particle density.  Some sources of turbulence could also generate pressure bumps \citep{Johansen2009}, but this effect cannot be simply parameterized and generally  applied to any disk model. Thus, we conservatively consider turbulence as a source of additional diffusive stirring.

Previous works have included the effects of turbulence on SI clumping by applying two criteria \citep[e.g.,][]{Drazkowska2016, Schoonenberg2017, Stammler2019}.  First, a metallicity threshold for pure SI (previous versions of our Equation \ref{eq:Z_crit}) must be met.  Second, the midplane dust-to-gas ratio must exceed unity, $\epsilon > 1$.  The approximate nature of this $\epsilon$ requirement is shown in Figure \ref{fig:H_p_epsilon}.

We offer an improvement on this previous approach in two ways.  First, a single criterion to consider for SI clumping instead of two, even with additional turbulence.  Second, we apply our more precise results for the $\epsilon_{\rm crit}$ needed to trigger strong clumping for different $\uptau_{\rm s}$.

As with $Z_{\rm crit}$, we use a piecewise quadratic function to fit $\log \epsilon_{\rm crit}$ as a function of $\log \uptau_{\rm s}$.  We obtain (again by eye)
\begin{equation}\label{eq:ep_crit}
  \log \epsilon_{\rm crit} \simeq A' (\log{\uptau_{\rm s}})^2 + B' \log{\uptau_{\rm s}} + C',
\end{equation}
with
\begin{align}
  \begin{cases}
    A'=0,\ B'=0,\ C'=\log{(2.5)} & \text{if } \uptau_{\rm s} < 0.015 \\
    A'=0.48,\ B'=0.87,\ C'=-0.11 & \text{if } \uptau_{\rm s} > 0.015
  \nonumber \end{cases}.
\end{align}
This fit is overplotted on the simulation results in Figure \ref{fig:H_p_epsilon}.

We assume that $\epsilon_{\rm crit}$ applies with additional sources of turbulence.  This assumption is roughly justified by the forced turbulence simulations of \citet{Gole2020}.  In the presence of both SI and additional turbulence, the particle scale-height is 
\begin{equation}\label{eq:Hpboth}
   H_{\rm p} = \sqrt{H_{{\rm p}, \eta}^2 + H_{{\rm p}, \alpha}^2} \, .
\end{equation}
We add these effects in quadrature, for simplicity, with $H_{{\rm p}, \eta} \equiv h_\eta \eta r$, as measured in our SI simulations, and $H_{\rm p, \alpha}/H = \sqrt{\alpha/(\alpha + \uptau_{\rm s})}$ for additional turbulence (e.g.\ \citealp{Youdin2007}).   We use $\alpha$ as the standard dimensionless measure of turbulent diffusivity, $\alpha H^2 \Omega$.

The midplane dust-gas ratio can thus be written as
\begin{equation}\label{eq:ep_alpha}
  \begin{aligned}
    \epsilon_\alpha &= \frac{\Sigma_{\rm p}}{\Sigma_{\rm g}} \frac{H}{H_{\rm p}} \\
    &= \frac{Z}{\sqrt{h_\eta^2 \Pi^2  + \alpha/(\alpha + \uptau_{\rm s})}}
  \end{aligned}
\end{equation}
The clumping condition $\epsilon_\alpha > \epsilon_{\rm crit}$, can be expressed as a condition on the particle metallicity as
\begin{equation}\label{eq:Z_alpha}
    \begin{aligned}
  Z > Z_{\rm crit, \alpha} &\equiv \epsilon_{\rm crit}(\uptau_{\rm s}) \sqrt{ h_\eta^2 \Pi^2 + \frac{\alpha}{(\alpha + \uptau_{\rm s}) }} \\
  &\simeq \epsilon_{\rm crit}(\uptau_{\rm s}) \sqrt{ \left( \frac{ \Pi}{5}\right)^2 + \frac{\alpha}{(\alpha + \uptau_{\rm s}) }} \, .
  \end{aligned}
\end{equation}

To apply Equation \ref{eq:Z_alpha}, we consider local disk values of $Z$, $\Pi$, $\alpha$ and $\uptau_{\rm s}$ and use Equation \ref{eq:ep_crit} for  $\epsilon_{\rm crit} (\uptau_{\rm s})$.  The final step of  Equation \ref{eq:Z_alpha} approximates $h_\eta \simeq 0.2$ which is sufficiently accurate, considering uncertainties in simulations and real disks.  Used in this way,  Equation \ref{eq:Z_alpha} is a single criterion for SI clumping that is more consistent with simulations than previous approaches.

While consistency of our $\epsilon_{\rm crit}$ with \citet{Gole2020} was already noted as a motivation, we can further test for consistency by applying Equation \ref{eq:Z_alpha} to the $\alpha$ values of their simulations.  The \citet{Gole2020} simulations fixed $\uptau_{\rm s} = 0.3$, $\Pi = 0.05$ and $Z = 0.02$.  For $\alpha = 10^{-3.5}$, they found strong clumping, consistent with Equation \ref{eq:Z_alpha} which gives $Z_{\rm crit, \alpha} = 0.013$ and thus $Z > Z_{\rm crit, \alpha}$. For the $\alpha = 10^{-3}$ case, \citet{Gole2020} found only very weak clumping, which is again consistent with our prediction of  $Z_{\rm crit, \alpha} = 0.022$, and thus $Z < Z_{\rm crit, \alpha}$, marginally.  More forced turbulence simulations with other parameters could further test and refine this clumping criterion.

%%%%%%%%%%%%%%%%%%%%%%%%%%%%%%%%%%%%%%%%%%%%%%%%%%%%%%%%%%%%
\subsection{Diffusion and Turbulence}
\label{subsec:diff_turb}

Since particle concentration and settling compete with turbulent mixing, we measure the radial and vertical diffusion coefficients in usual ways.  The radial diffusion coefficient is extracted from the best fit to
\begin{equation}\label{eq:D_px}
  D_{\mathrm{p}, x} \equiv \frac{1}{2}\fracp{\sigma_{x}^2}{t},
\end{equation}
as in \citet[][]{Johansen2007}, where $\sigma_{x}^2$ denotes the variance of the radial particle positions, $x_i(t)- x_i(t_0)$.  We take the initial position at $t_0 = t_{\rm sedi-tr}$, after initial sedimentation.  Due to vertical gravity, particles do not diffuse freely in $z$.  Thus, we measure the vertical diffusion coefficient by balancing the particle settling and diffusion timescales \citep{Dubrulle1995, Youdin2007}
\begin{equation}\label{eq:D_pz}
  D_{\mathrm{p}, z} = \frac{H_{\rm p}^2}{t_{\rm sett}} \equiv H_{\rm p}^2 \Omega_0 \uptau_{\rm s},
\end{equation}
which is valid for $H_{\rm p} \ll H$.

Figure \ref{fig:D_p} shows both diffusion coefficients as a function of $\uptau_{\rm s}$ for the $Z$ values simulated (and Table \ref{tab:prop_2D} lists the coefficients).  The maximum value of the diffusion coefficients (at $\uptau_{\rm s} =1$) is roughly $\alpha \equiv D_{\rm p}/( c_{\rm s} H) \sim 10^{-4}$, which is expected from the basic scales of the SI, as $(\eta r)^2 \Omega = 2.5 \times 10^{-4} c_{\rm s} H$ in our simulations.  The trend of diffusion increasing with $\uptau_{\rm s}$ can be understood in two ways.  First, a more vigorous SI at larger $\uptau_{\rm s}$ is expected from faster linear, unstratified growth rates \citepalias{Youdin2005}.  Also, since $H_{\rm p}$ varies only weakly with $\uptau_{\rm s}$, stronger diffusion is needed to stir larger particles (as indicated by Equation \ref{eq:D_pz}).  However, $H_{\rm p}$ does vary somewhat with $\uptau_{\rm s}$.  Correspondingly the scaling of $D_{\rm p}$ with $\uptau_{\rm s}$ is steeper than linear from $0.01$ to $0.1$ and shallower outside this range.

While $D_{\mathrm{p}, x}$ and $D_{\mathrm{p}, z}$, show a similar $\uptau_{\rm s}$ dependence, the measured values of $D_{\mathrm{p}, z}$ are consistently smaller, by a factor of $\sim 2$ -- $10$.  While anisotropic diffusion is a real possibility, we caution against over-interpreting this result.  Particle stirring is not precisely described by a diffusive random walk.  In the vertical direction, density structures on large scales -- up to $\sim H_{\rm p}$, see Figure \ref{fig:rhop_xz_tauZ} -- are inconsistent with a random walk with small steps $\ll H_{\rm p}$.   In the radial direction, weak particle clumping during the pre-clumping phase biases diffusion measurements.  The slower drift of particles in  weak clumps makes a non-diffusive contribution to the measured variance $\sigma_x^2$.  This effect explains the counter-intuitive result that measured $D_{{\rm p},x}$ can be larger for cases that eventually produce strong clumping (e.g. $\uptau_{\rm s} = 0.02, 0.03$ and $0.3$ runs).  For $D_{\mathrm{p}, z}$, the effect of (early, weak) clumping is opposite  at least for $\uptau_{\rm s} \leq 0.1$.  Since clumps tend to reside close to the midplane, $D_{\mathrm{p}, z}$ tends to be smaller in the clumping cases.  For $\uptau_{\rm s} \geq 0.3$ the corrugation of the dust layer (which is distinctly non-diffusive as noted above) changes this trend.  Despite these complexities, the similarity of $D_{\mathrm{p}, x}$ and $D_{\mathrm{p}, z}$ values suggests that particle mixing is at least approximately diffusive in nature.

As with the related $H_{\rm p}$ values, the diffusion coefficients do not give a clear indication of which cases will or will not clump.  The complex balance between clumping and diffusion in nonlinear SI is unfortunately not elucidated in obvious trends of $D_{\rm p}$.  The diffusion values are useful in another way, indicating the strength of other turbulence needed to interfere with SI clumping.  This effect is included in the revised clumping boundary with turbulence,  Equation \ref{eq:Z_alpha}.

%%%%%%%%%%%%%%%%%%%%%%%%%%%%%%%%%%%%%%%%%%%%%%%%%%%%%%%%%%%%
\subsection{Comparison to Linear Theory}
\label{subsec:comp_s}

We examine whether linear, unstratified SI growth rates (\citetalias{Youdin2005}; \citealp{YJ2007}) are a useful predictor of clumping in stratified non-linear simulations.  We find that strong clumping does not correlate with linear growth rates.  Perhaps a strong connection should not be expected as there are three significant distinctions: linear vs.\ non-linear, stratified vs.\ unstratified, and particle clumping vs. growth (e.g., of velocities more than particle densities).

Previous works have already demonstrated a disconnection between linear growth rates and clumping.  For unstratified SI, non-linear clumping is weak for $\uptau_{\rm s} = 0.1$, $\epsilon = 3$ even though the linear growth rate is faster than the $\uptau_{\rm s} = 1$, $\epsilon = 0.2$ case that produces strong clumping \citep{Johansen2007}.   For related vertical settling instabilities \citep{Squire2018}, simulations show weak particle clumping even for fast linear growth \citep{Krapp2020}. 

More studies of these connections are warranted, as linear growth is frequently used as a guide to clumping when simulations are unavailable.  The recent development of a linear theory for stratified SI \citep{Lin2021} is a significant advance, and comparison of simulations to this theory is a goal of future work.

%%%%%%%%%%%%%%%%%%%%%%%%%%%%%%%%%%%%%%%%%%%%%%%%%%%%%%%%%%%%
\begin{figure*}[htbp!]
  \centering
  \includegraphics[width=\linewidth]{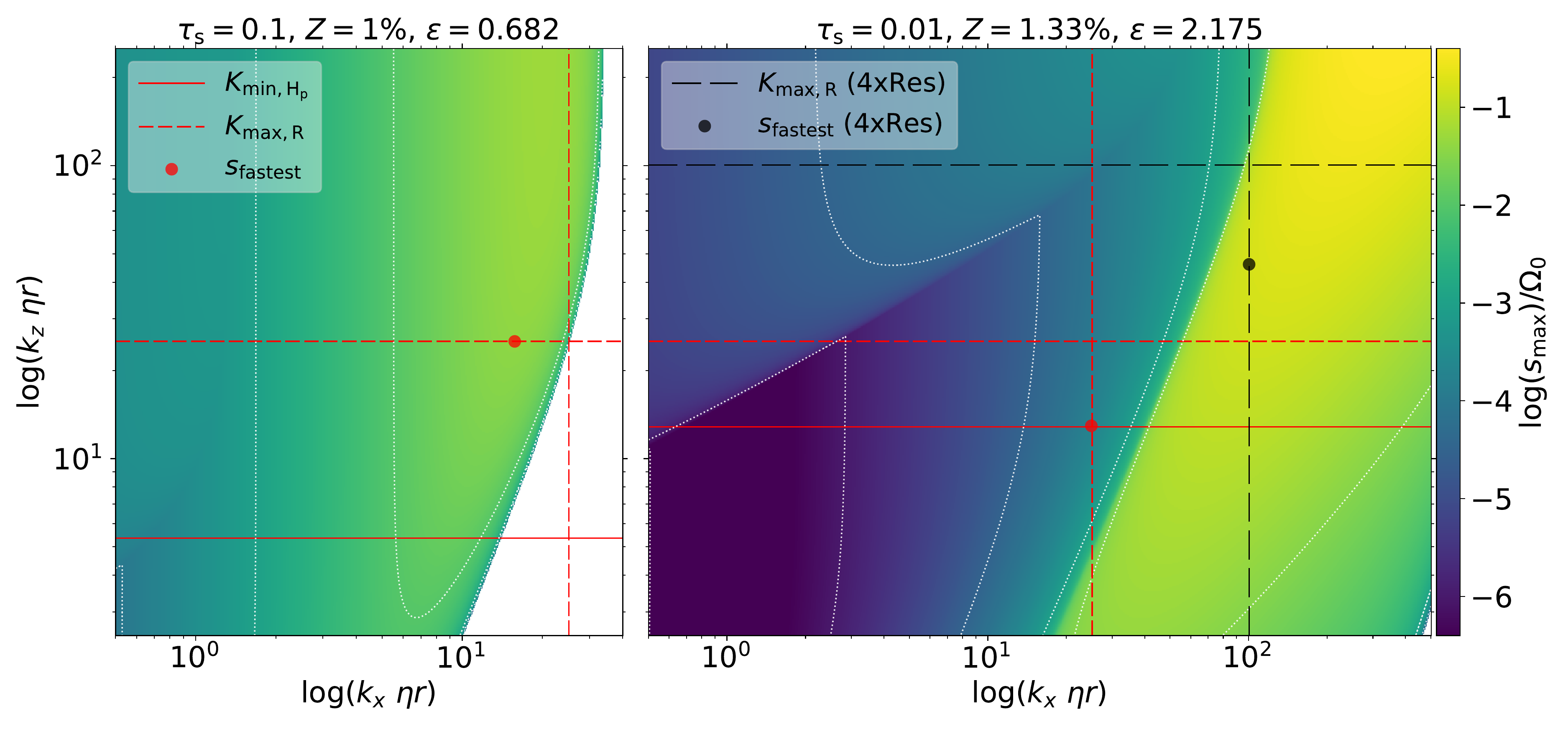}
  \caption{Maps of streaming instability growth rates as a function of the radial and vertical wavenumbers, for the parameters of Run \texttt{Z1t10} (\textit{left}) and Run \texttt{Z1.33t1} (\textit{right}).  Simulated $\epsilon$ values are used to compute standard unstratified growth rates.  Wavenumbers that meet the resolution requirements of our standard resolution runs must lie below and to the left of the \textit{dashed red lines}.  Wavenumbers must also lie above the \emph{solid red line} to fit in the dust layer.  (See text for details.)  The fastest growing modes that meet these requirements are indicated by \textit{red dots}.  A run with 4 times the resolution (\emph{right panel only}) would have the resolution requirements set by the \textit{black long dashed lines} (and the size requirement set by the same solid red line), which implies a much faster resolved growth rate, at the location of the \textit{black dot}.
  \label{fig:eg_s_max}}
\end{figure*}

\begin{figure*}[htbp!]
  \centering
  \includegraphics[width=\linewidth]{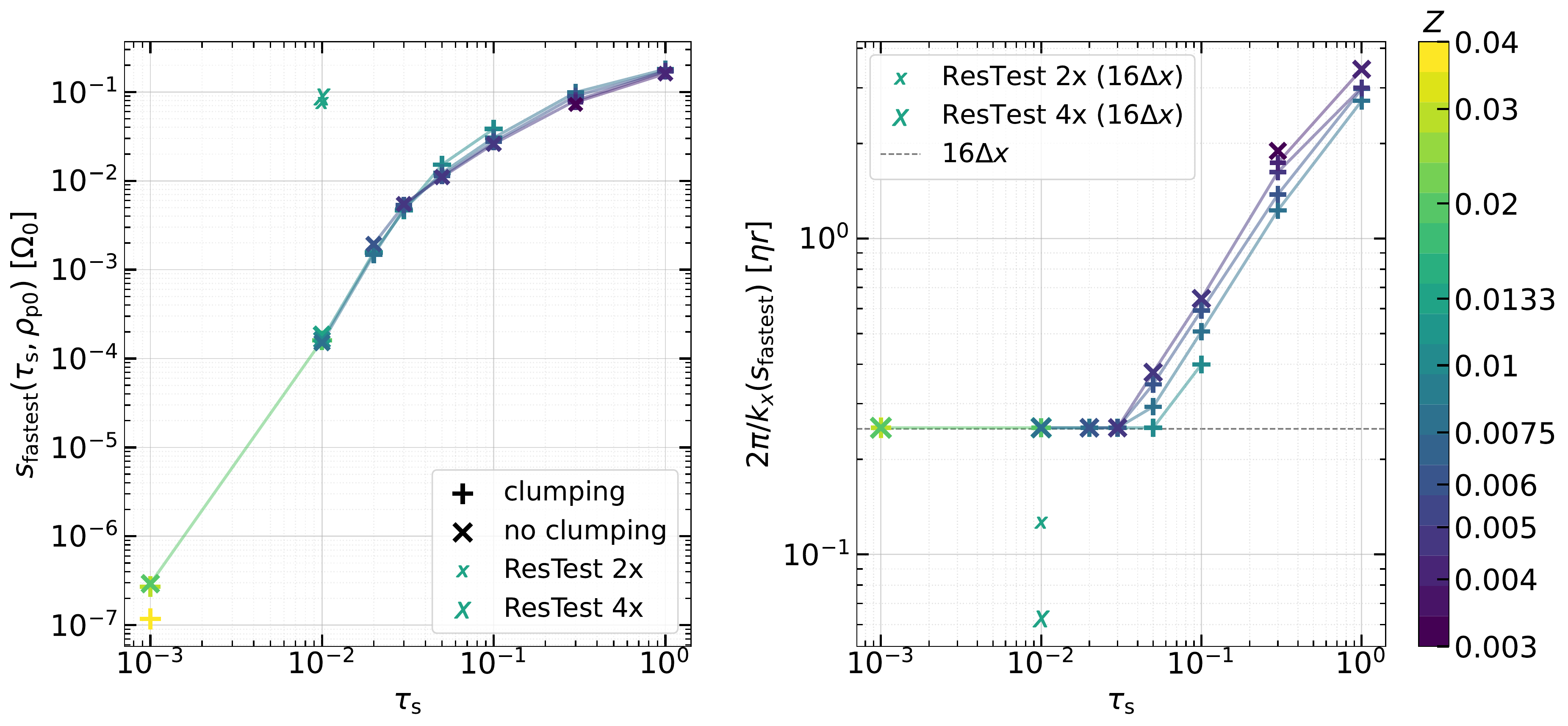}
  \caption{(\textit{Left:}) The maximum growth rates of the unstratified streaming instability that satisfy the requirements illustrated in Figure \ref{fig:eg_s_max}.  (\textit{Right:}) The corresponding radial wavelengths.  Symbols, lines, and their colors have the same meaning as Figure \ref{fig:H_p_epsilon}.  These growth rates are nearly identical for clumping and non-clumping cases, and thus a poor predictor of clumping.
  The higher resolution ``ResTest'' runs nominally resolve fast growth, but still do not produce clumping (see Section \ref{subsubsec:num_res}).
  The \emph{black dotted line (right panel)} indicates the resolution limit at standard resolution.  \label{fig:s_max}}
\end{figure*}

%%%%%%%%%%%%%%%%%%%%%%%%%%%%%%%%%%%%%%%%%%%%%%%%%%%%%%%%%%%%
\subsubsection{Resolution and size requirements}
\label{subsubsec:res_req}

To find the most relevant unstratified growth rates, we first compare the resolution of our simulations to the wavelengths of linear (unstratified) SI modes.  We use dimensionless wavenumbers $K = k \eta r$ and a grid cell size $\Delta x = \eta r/N_\eta$, where $N_\eta = 64$ in our simulations.  A mode is deemed nominally resolved with $R_{\rm min} = 16$ cells per wavelength,
\footnote{Higher resolution may be needed for numerical convergence of slower growth rates \citep{Bai2010}.  A uniform resolution requirement is not only simpler, but ``conservative'' in the sense that our surprising findings (i.e.\ clumping with nominally slow linear growth growth rates) would be reinforced if growth rates were even slower due to stricter resolution requirements.}
i.e.\ if the wavenumber
\begin{equation}\label{eq:K_maxR}
  \displaystyle K < K_{\rm max, R} = 2 \pi \frac{N_\eta}{R_{\rm min}} \simeq  25 \left( \frac{N_\eta}{64} \right).
\end{equation}
In practice, we apply Equation \ref{eq:K_maxR} individually to both $K_x$ and $K_z$, instead of the total $K$.

Also, the vertical wavelength should be small enough for modes to fit within the particle layer.   Specifically, we require that at least half a wavelength fit within the particle layer width, $2 H_{\rm p}$, i.e., 
\begin{equation}\label{eq:K_min_H_p}
  K_z > K_{\rm min, H_{\rm p}} = \frac{\pi}{2 H_{\rm p}} \eta r = \frac{\pi}{2} \frac{1}{0.2 f_H} \simeq \frac{8}{f_H},
\end{equation}
where $f_H \equiv H_{\rm p} / 0.2 \eta r \sim 1$ (see Figure \ref{fig:H_p_epsilon}).  This cutoff is the most generous (in terms of number of linear modes) that we could justify.  When applying Equation \ref{eq:K_min_H_p} to a specific run, we use the measured $H_{\rm p}$, not an  approximation of $f_H$.

Stratified simulations should also resolve the particle scale height, independent of any linear theory.  The dust sublayer width in our simulations is well resolved with
\begin{equation}
  \frac{2 H_{\rm p}}{\Delta x} = \frac{2}{\Delta x}\ 0.2 f_H \eta r \simeq 26 f_H \left(\frac{N_\eta}{64}\right) >  R_{\rm min}.
\end{equation}
Since this requirement is always satisfied, it doesn't affect our analysis.

%%%%%%%%%%%%%%%%%%%%%%%%%%%%%%%%%%%%%%%%%%%%%%%%%%%%%%%%%%%%
\subsubsection{Linear Growth Rates vs.\ Clumping}
\label{subsubsec:s_vs_clumping}
To compare unstratified linear theory to our stratified simulations, we use the midplane $\epsilon$ values as measured in Section \ref{subsec:rhop} to calculate linear SI growth rates.\footnote{We include gas compressibility (a weak effect) in the eigenvalue analysis as in \citet{YJ2007} and report the fastest growing of the 8 modes.}
Figure \ref{fig:eg_s_max} maps SI growth rates versus wavenumber for two example cases, corresponding to the parameters of runs \texttt{Z1t10} and \texttt{Z1.33t1}.  These plots show that SI growth is fastest along contours that roughly follow $K_z \propto K_x^2$ at smaller $K$ and constant $K_x$ at larger $K$ (\citetalias{Youdin2005}; \citealp{Squire2018}).  Figure \ref{fig:eg_s_max} also shows the wavelength restrictions imposed by Equations \ref{eq:K_maxR} and \ref{eq:K_min_H_p}, and locates the fastest growth rate, $s_{\rm fastest}$, that meets these requirements.

The left panel of Figure \ref{fig:s_max} presents these linear growth rates, $s_{\rm fastest}$, for all our runs.  The main takeaway from this plot is that there is no significant difference in the growth rates of clumping and non-clumping runs.  This result is surprising, especially since the clumping cases have higher $\epsilon$ values (see Figure \ref{fig:H_p_epsilon}).  Even though higher $\epsilon$ values can produce faster growth of linear SI, this faster growth occurs at smaller wavelengths.  Moreover, the changes to $\epsilon$  across the clumping boundary are relatively modest.  These factors help explain why linear growth rates -- at fixed resolution -- change so little when comparing parameters of clumping and non-clumping simulations.  We emphasize that we are not demonstrating any inconsistency in either linear, unstratified SI growth or non-linear, stratified SI clumping.  Rather we are showing that these two problems are not as closely related as some might assume.

The right panel of Figure \ref{fig:s_max} plots the radial wavelengths of modes growing at the rate $s_{\rm fastest}$.  For $\uptau_{\rm s} < 0.05$ modes are only marginally resolved in radius, i.e.\ the fastest growth is  at the imposed radial resolution requirement.\footnote{It is impossible to resolve the fastest growing modes in $z$ as the fastest growth is for $K_z \rightarrow \infty$, absent diffusive effects \citepalias{Youdin2005}.}  
Furthermore, the slope of $s_{\rm fastest}$ versus $\uptau_{\rm s}$ is also much steeper for $\uptau_{\rm s} < 0.05$, as shown in the left panel.  This transition can be understood in terms of the properties of linear SI as we now explain.  While the following explanation is somewhat detailed, it is relevant to understanding the desired resolution of SI simulations. 

The fastest growth of linear SI occurs for $K_x = K_{x, \infty} \simeq f_\epsilon/(2\uptau_{\rm s})$ and $K_z \gtrsim K_{x, \infty}$, i.e., at smaller wavelengths for smaller $\uptau_{\rm s}$ \citepalias{Youdin2005}.  The numerical factor $f_\epsilon \rightarrow 1$ for small epsilon \citep[as explained by][]{Squire2018} and increases with $\epsilon$ (see \citealp{Lin2017} for a large $\epsilon$ limit).  We find  $f_\epsilon \lesssim 10$ in our runs with $\epsilon \sim 1$. 

Applying this  $K_{x, \infty}$ to the resolution condition of Equation \ref{eq:K_maxR}, the fastest growth should be resolved for
\begin{equation}\label{eq:taures}
 \uptau_{\rm s} \gtrsim  0.1 \frac{f_\epsilon}{5} \left( \frac{64}{N_\eta} \right).
\end{equation}
This condition agrees, within a factor $\sim$2, with the transition noted above.  The slight difference is explained by the fact that for $K_x < K_{x, \infty}$ relatively fast growth still occurs along a $K_z \propto K_x^2$ contour (\citetalias{Youdin2005}; \citealp{Squire2018}).  However, only slightly larger radial wavelengths are allowed on this contour, due to the vertical size constraint (Equation \ref{eq:K_min_H_p}) and the nonlinear dependences of $K_z \propto K_x^2$ along the contour.  
This analysis of $s_{\rm fastest}$ and the corresponding $K_x$ agrees with the growth rate maps (similar to Figure \ref{fig:eg_s_max}) shown at \href{https://rixinli.com/tauZmap.html}{https://rixinli.com/tauZmap.html}.

Equation \ref{eq:taures} thus provides a theoretical resolution requirement for SI simulations, i.e.\ a constraint on $N_\eta$.  However, this condition assumes a connection between linear, unstratified growth and non-linear, stratified simulations.  As the tests presented here cast doubt on the strength of this connection, empirical resolution tests (as in \citealp{Yang2017}) remain crucial, and are considered below (Section \ref{subsubsec:num_res}).

%%%%%%%%%%%%%%%%%%%%%%%%%%%%%%%%%%%%%%%%%%%%%%%%%%%%%%%%%%%%
\subsubsection{Linear growth with turbulence}
\label{subsubsec:s_damped}
To more precisely compare with our non-linear simulations, we can add the effects of diffusive stabilization.  Specifically, growth rates can be damped as \citepalias{Youdin2005}
\begin{equation}\label{eq:s_damped}
    s_{\rm damped} = s_{\rm max} -D_{{\rm p},x} k_x^2 - D_{{\rm p},z} k_z^2
\end{equation}
using the measured diffusion coefficients.  Unfortunately, with this damping there are no positive relevant growth rates for any simulations.  As above, long wavelengths are disallowed by the size requirement of Equation \ref{eq:K_min_H_p}. Even if the radial diffusion coefficient $D_{{\rm p},x}$ is replaced by the smaller $D_{{\rm p},z}$, the $s_{\rm damped}$ remain negative.

This strong inconsistency between  stratified simulations and linear unstratified theory with turbulent diffusion is discouraging, but an important caution about the over-interpretation of linear theory out of context.

%%%%%%%%%%%%%%%%%%%%%%%%%%%%%%%%%%%%%%%%%%%%%%%%%%%%%%%%%%%%
\begin{figure}[tbh!]
  \centering
  \includegraphics[width=\linewidth]{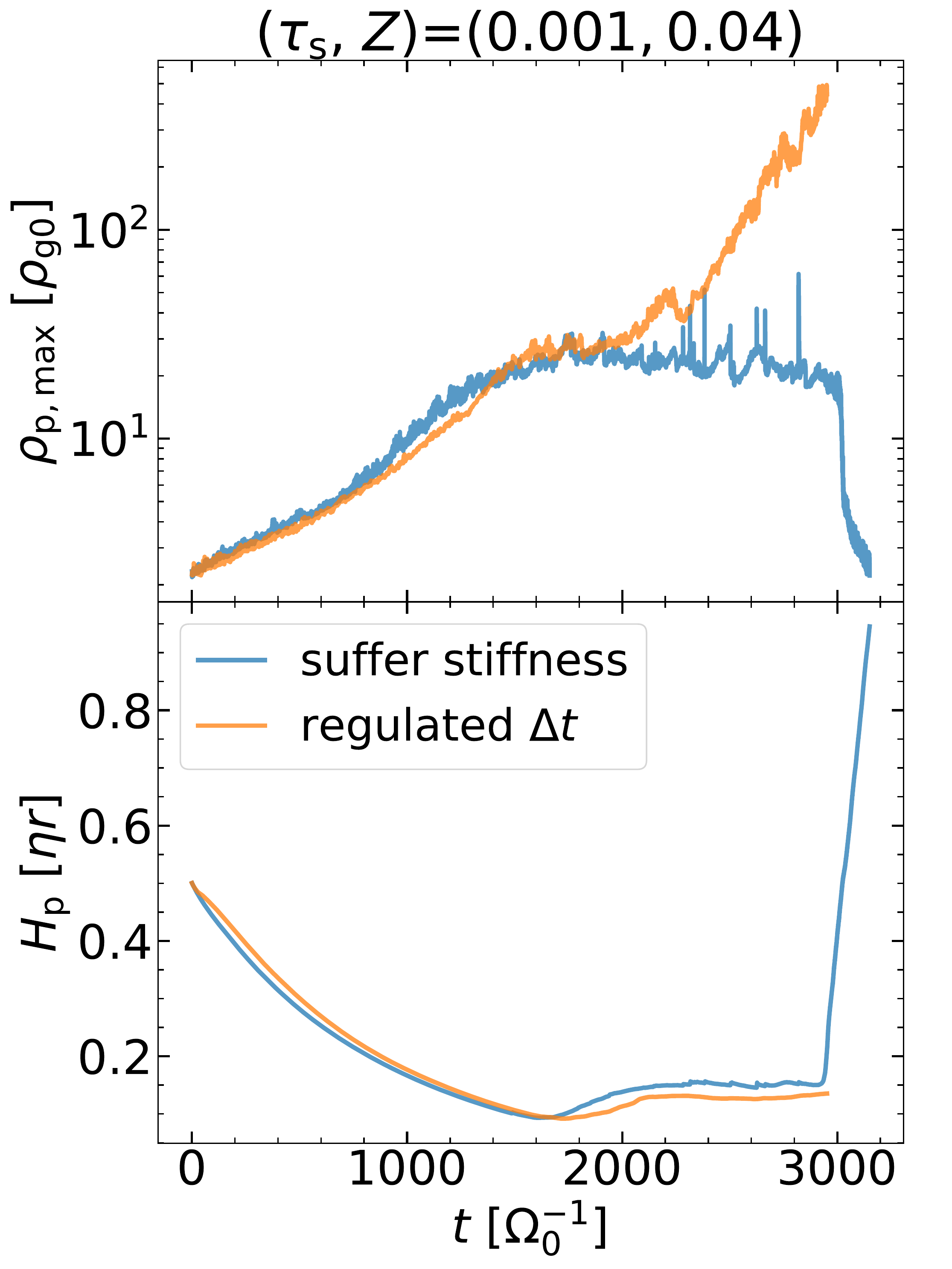}
  \caption{Maximum particle density and particle scale height as a function of time for Run \texttt{Z4t01}, comparing the results before and after adopting regulated timestep (see Equation \ref{eq:dt_stiff}), which resolves the stiffness issue.  \label{fig:stiffness}}
\end{figure}

\begin{figure*}[htbp!]
  \centering
  \includegraphics[width=\linewidth]{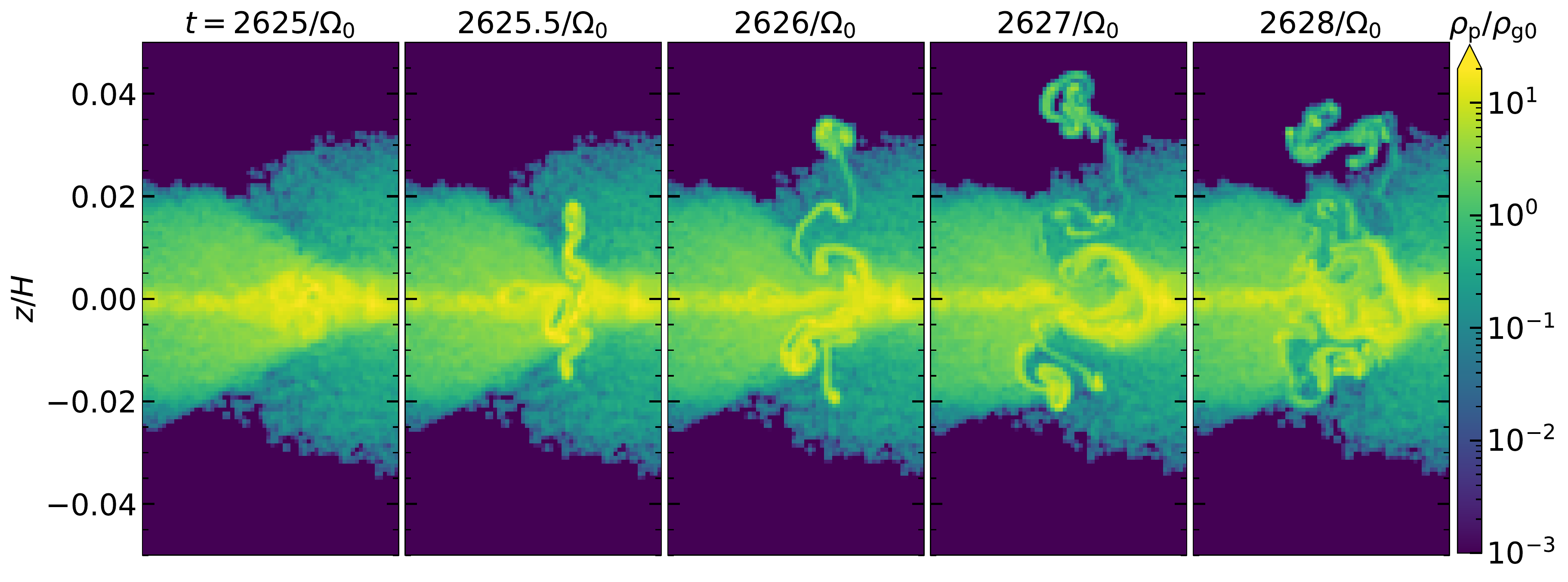}
  \caption{An example of numerical instability in the particle layer caused by high stiffness.  The ``explosion'' that reinflates the particle layer is unphysical, and preventable with stricter timestep controls (see Figure \ref{fig:stiffness}).
  \label{fig:stiff_explosion}}
\end{figure*}
%%%%%%%%%%%%%%%%%%%%%%%%%%%%%%%%%%%%%%%%%%%%%%%%%%%%%%%%%%%%

%%%%%%%%%%%%%%%%%%%%%%%%%%%%%%%%%%%%%%%%%%%%%%%%%%%%%%%%%%%%%%%%%%%%%%%%%%%%%%%
\section{Discussion}
\label{sec:discussion}

To better understand the robustness of our numerical results, we analyze relevant numerical issues in Section \ref{subsec:numerical} and compare to previous works on the clumping boundary for SI in Section \ref{subsec:comp_pre}.  
A common theme is that improved numerics in SI simulations tend to produce stronger clumping, and the rare exceptions to the rule are cases of weak clumping.  These trends suggest, reassuringly, that perfect solutions would have similar, or more, clumping than achievable simulations.

%%%%%%%%%%%%%%%%%%%%%%%%%%%%%%%%%%%%%%%%%%%%%%%%%%%%%%%%%%%%
\subsection{Numerical Effects}
\label{subsec:numerical}

%%%%%%%%%%%%%%%%%%%%%%%%%%%%%%%%%%%%%%%%%%%%%%%%%%%%%%%%%%%%
\subsubsection{Stiffness}
\label{subsubsec:stiffness}

The coupled particle-gas system becomes stiff when the particle size is small and especially when the particle-to-gas density ratio is high.  \citet{Bai2010} defined a stiffness parameter
\begin{equation}
  \chi \equiv \frac{\epsilon \Delta t_{\rm CFL}}{\text{max}(\uptau_{\rm s}, \Delta t_{\rm CFL})},
\end{equation}
where $\Delta t_{\rm CFL} \sim C_0 \Delta x / c_{\rm s}$ is the time step in \texttt{ATHENA} that satisfies the Courant–Friedrichs–Lewy (CFL) stability condition based on wavespeeds (i.e., of sound waves), where $C_0$ is a fixed CFL number of $0.4$ (see also \citealt{Stone2008}).  When $\chi \gtrsim 3$--$5$, numerical instabilities emerge and may result in unphysically large growth of particle and gas velocities, which inhibit clumping.  For our simulations, $\Delta t_{\rm CFL} \simeq 3\times10^{-4} / \Omega_0$ at the standard resolution.  Thus, $\chi = \epsilon \Delta t_{\rm CFL} / \uptau_{\rm s}$, indicating our runs with $\uptau_{\rm s} = 0.001$ are easily unstable when $\epsilon \gtrsim 20$, much smaller than $\rho_{\rm R32}$.

\citet{Bai2010} implemented a technique to enforce $\chi < 3$
\footnote{In the public version of \texttt{ATHENA}, $\chi$ is further enforced to be $< 1$.  A similar technique is also adopted in the \texttt{Pencil} Code before \citet{Yang2016}, who implemented a new algorithm for stiff mutual drag interaction.}
by effectively increasing the value of $\uptau_{\rm s}$ for all local particles within relevant dense regions.  In addition, the predicted momentum feedback in the first step of the predictor-corrector scheme is reduced by max$(\chi, 1)$ to further improve the stability (see \citealp{Bai2010} for more details).  

While this technique is adequate for many applications, we noted some difficulties.  First, by changing stopping times and momentum feedback, the technique gives an inherently less accurate solution, which might be insufficient for detailed SI dynamics at high $\epsilon$.  Second, when using this technique, numerical instabilities still arose in stiff regimes, for example, for Run \texttt{Z4t01} when $\rho_{\rm p}/\rho_{\rm g} \gtrsim 20$ (see Figures \ref{fig:stiffness} and \ref{fig:stiff_explosion}).

Thus we adopted an alternate technique to handle stiffness, which is both more physically accurate and more computationally expensive.  Our brute force approach restricts the timestep to 
\begin{equation}\label{eq:dt_stiff}
  \Delta t = \text{min}\left(\Delta t_{\rm CFL}, \frac{\chi_{\rm th} \text{max}(\Delta t_{\rm CFL}, \uptau_{\rm s})}{\epsilon} \right),
\end{equation}
which ensures that the timestep will both satisfy the Courant condition and keep the stiffness parameter below $\chi_{\rm th} = 0.75$.  This conservative limit empirically secured numerical stability in all our runs.

Due to the extra computational expense of this approach, runs with high stiffness must be terminated after strong clumping beings.  Specifically, Run \texttt{Z4t01} ends at $t=2952/\Omega_0$.

%%%%%%%%%%%%%%%%%%%%%%%%%%%%%%%%%%%%%%%%%%%%%%%%%%%%%%%%%%%%
\begin{figure}[htbp!]
  \centering
  \includegraphics[width=\linewidth]{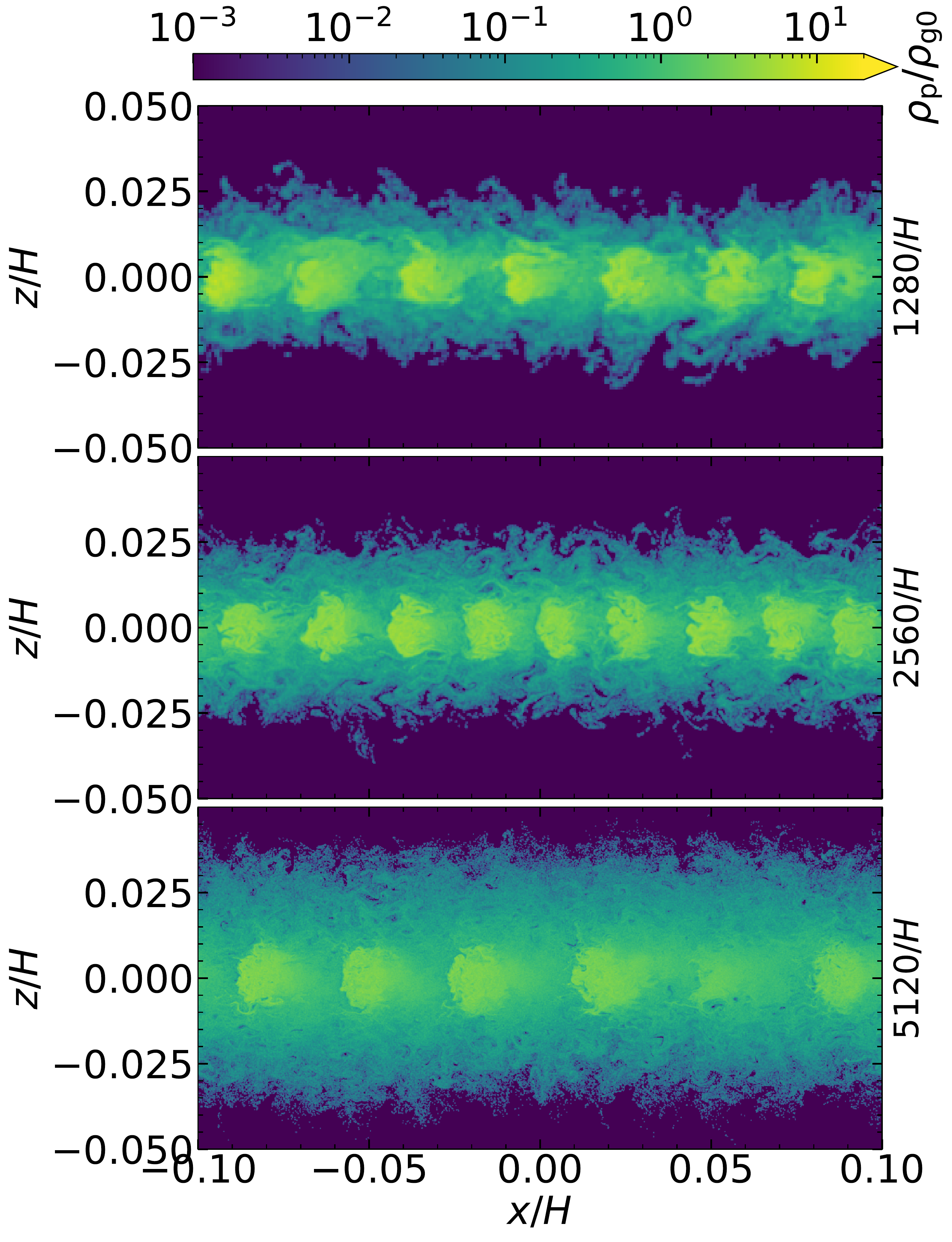}
  \caption{Final particle density for the resolution test with $Z = 1.33\%$ and $\uptau_{\rm s} = 0.01$. Resolution increases from top (standard) to bottom (4x higher).  This case has $Z$ just below the strong clumping boundary.  Higher resolution produces slightly weaker clumping and a thicker particle later.  The smaller box sizes of the higher resolution runs (images do not show the entire domain) contributes to these trends.  See Section \ref{subsubsec:num_res} for discussion.
  \label{fig:rhop_xz_resT}}
\end{figure}

\begin{figure}[htbp!]
  \centering
  \includegraphics[width=\linewidth]{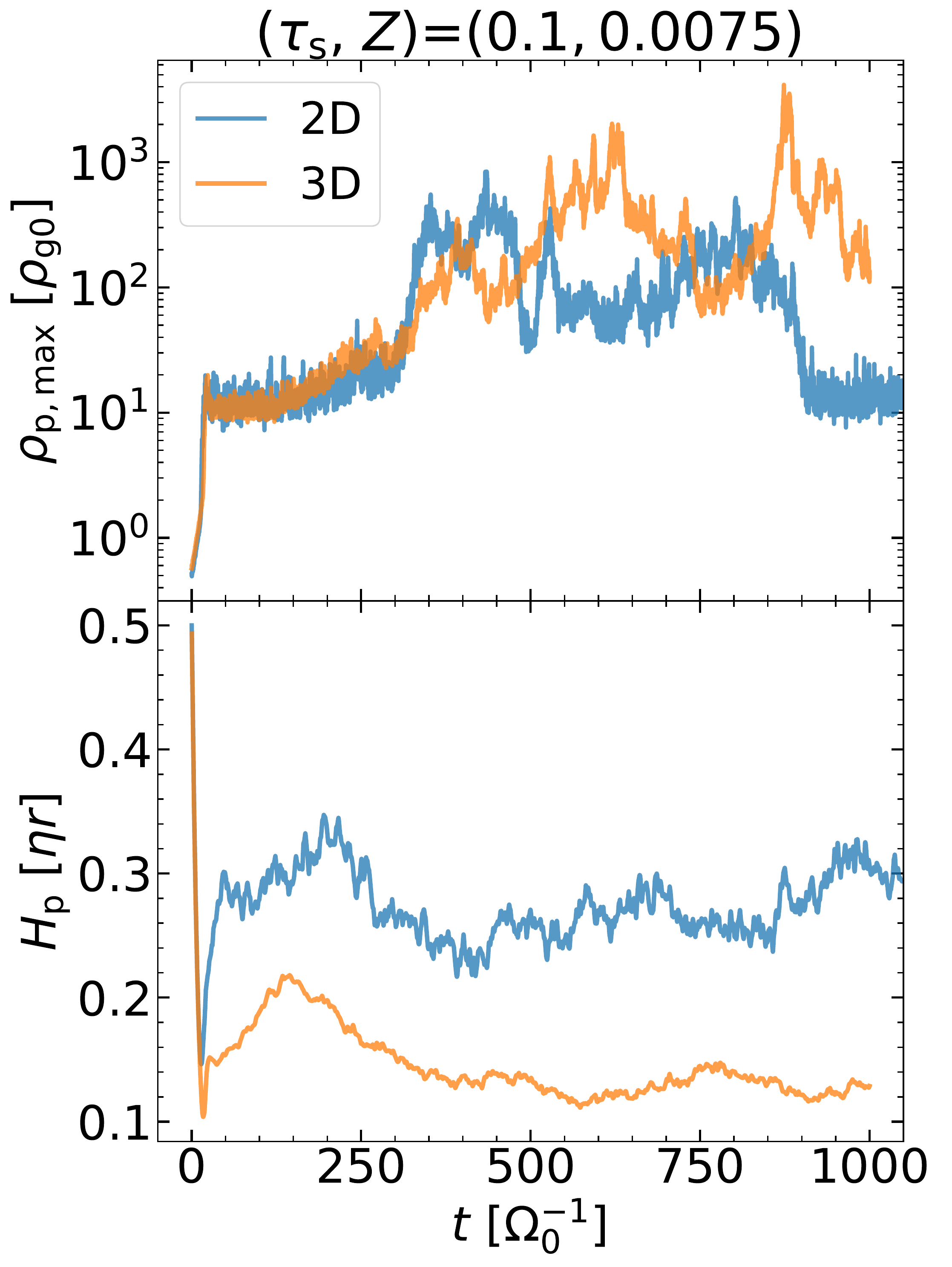}
  \caption{Comparison of 2D and 3D runs showing the evolution of maximum particle density and particle scale height for two runs (\texttt{Z0.75t10} and \texttt{3D-Z0.75t10}) with the same physical parameters, ($\uptau_{\rm s}, Z$) = ($0.1, 0.0075$).  The azimuthal direction provides another dimension for particles to concentrate, leading to smaller particle scale height and higher maximum particle density.  \label{fig:parStats_2D3D}}
\end{figure}

\begin{figure*}[htbp!]
  \centering
  \includegraphics[width=\linewidth]{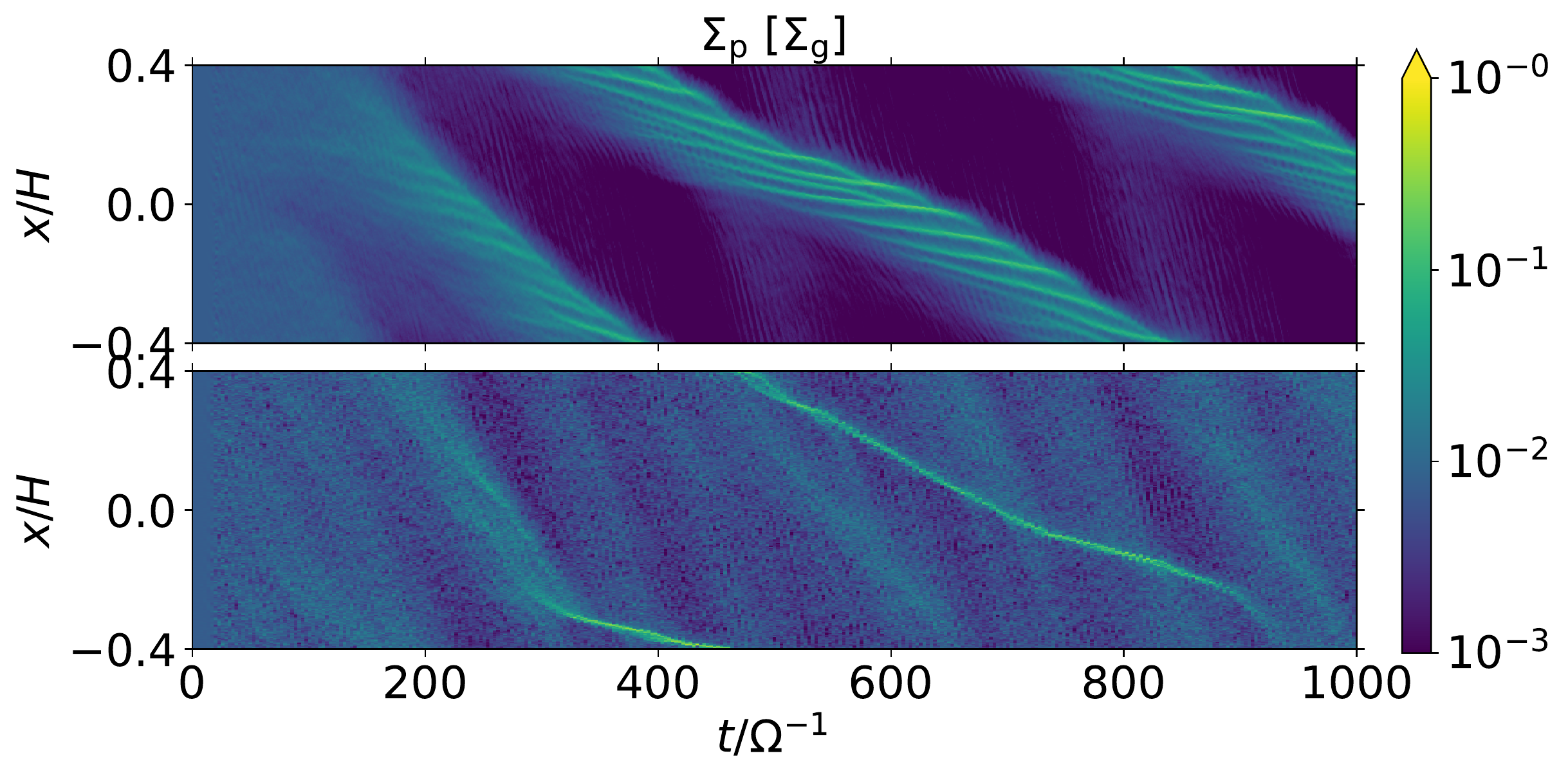}
  \caption{Comparison of the radial concentration of particles as a function of time between the 3D (\textit{upper}) and 2D (\textit{lower}) models shown in Figure \ref{fig:parStats_2D3D}. %\any{\emph{Would this be better showing only first 1000 / $\Omega$ for both?}}
  \label{fig:rhop_xt_2D3D}}
\end{figure*}

\begin{figure*}[htbp!]
  \centering
  \includegraphics[width=\linewidth]{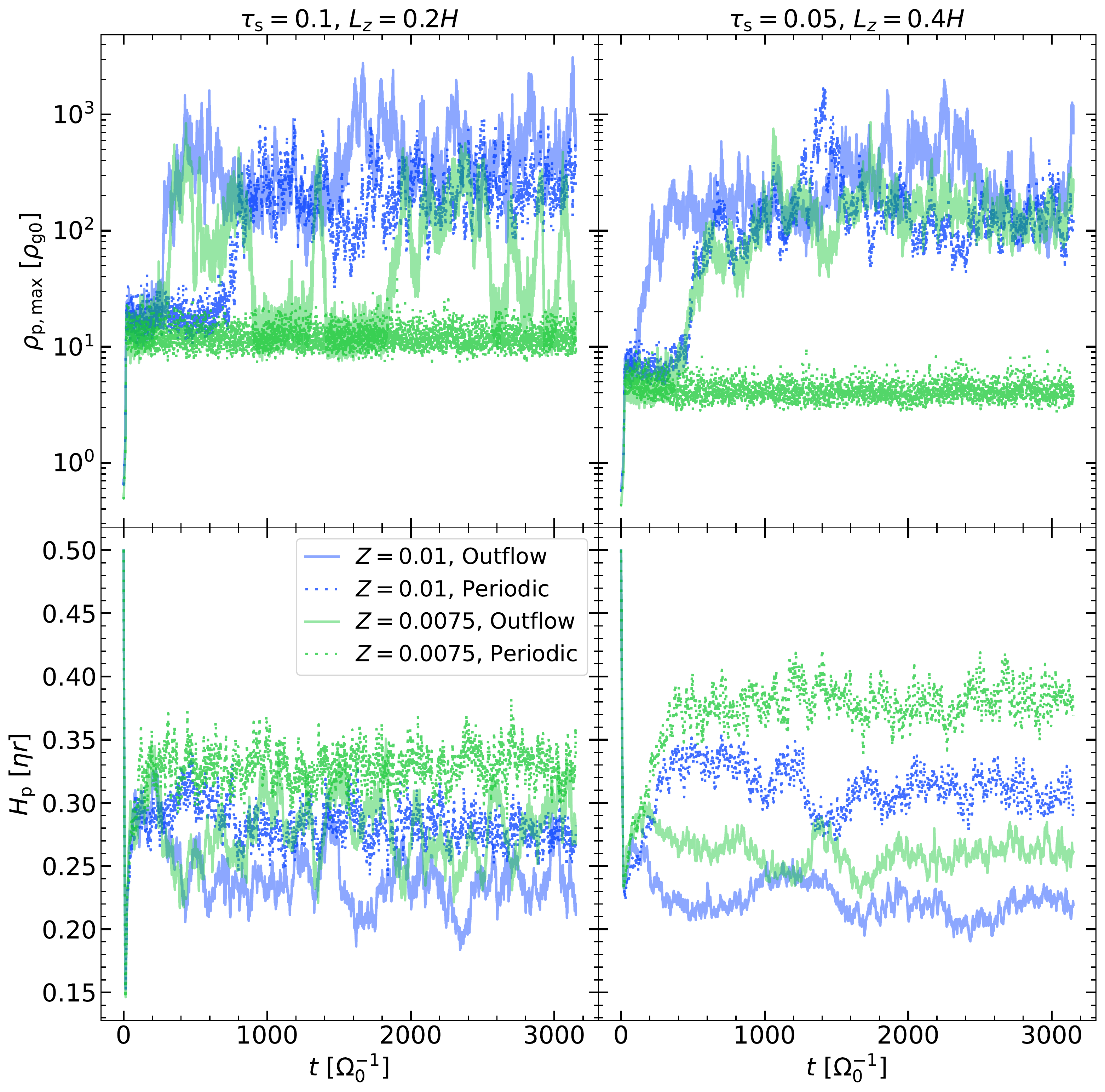}
  \caption{A comparison of outflow (\emph{solid curves}) )vs. periodic vBCs (\emph{dotted curves}) showing maximum particle density (\emph{top panels}) and particle scale height (\emph{bottom panels}) as a function of time for models with $\uptau_{\rm s} = 0.1 (\emph{left}), 0.05 (\emph{right})$ and $Z = 0.01, 0.0075$.  
  Our preferred outflow vBCs allow greater particle settling and promote particle clumping.  See Section \ref{subsec:boundary} for discussion.
  \label{fig:comp_ou_pe}}
\end{figure*}

\begin{figure}[htbp!]
  \centering
  \hspace{-2em} \makebox[\linewidth][c]{\includegraphics[width=1.1\linewidth]{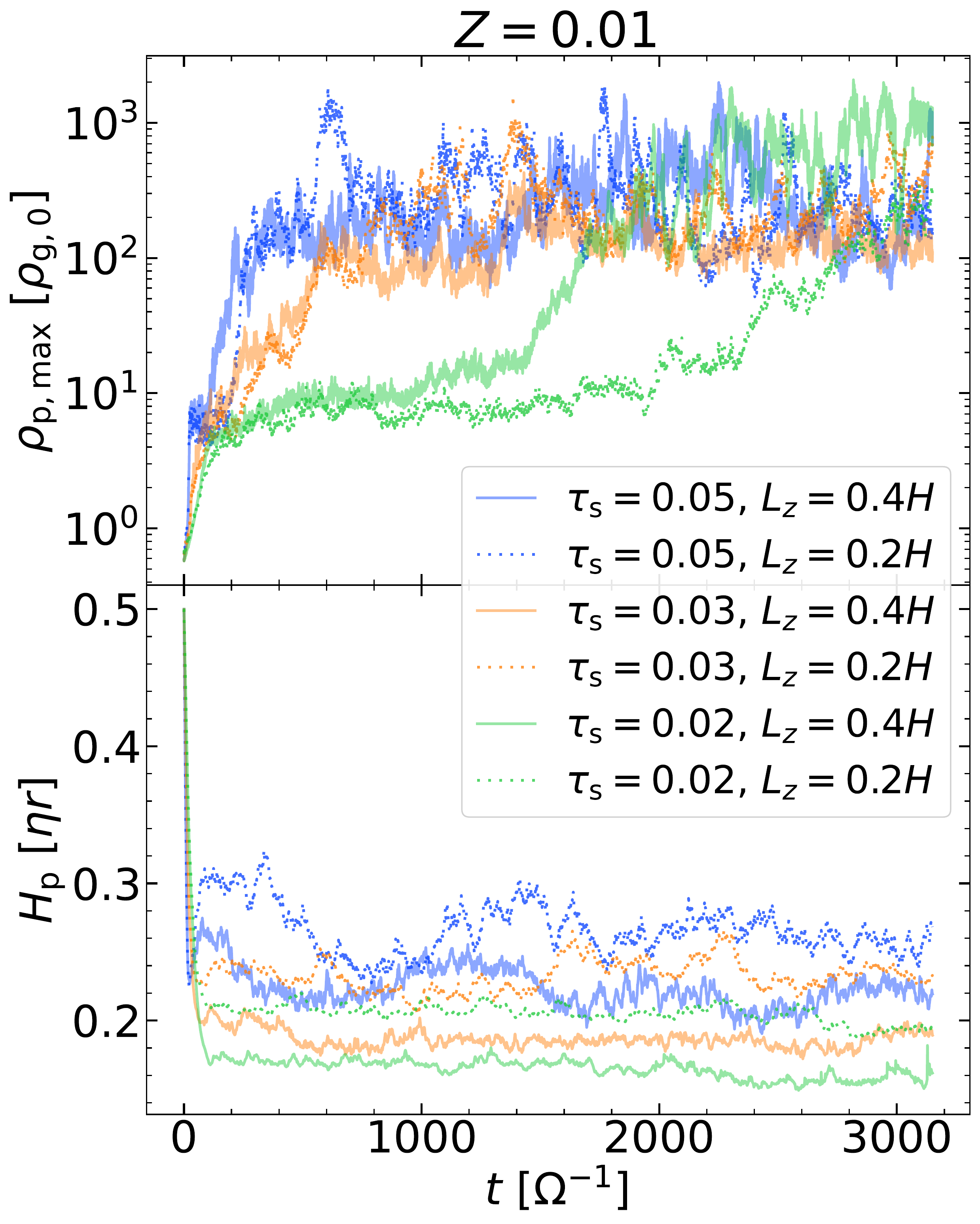}}
  \caption{Similar to Figure \ref{fig:comp_ou_pe}, but comparing our standard $L_Z=0.4H$ runs (\emph{solid curves}) to shorter box simulations with $L_Z=0.2H$ (\emph{dashed curves})  for models with $Z = 0.01$ and $\uptau_{\rm s} = 0.05, 0.03, 0.02$.  In the taller boxes there is more particle settling, and (for $\uptau_{\rm s} = 0.02$) there is more rapid and stronger particle clumping.
  \label{fig:comp_Lz}}
\end{figure}

%%%%%%%%%%%%%%%%%%%%%%%%%%%%%%%%%%%%%%%%%%%%%%%%%%%%%%%%%%%%
\subsubsection{Resolution}
\label{subsubsec:num_res}

Since perfect convergence of nonlinear SI simulations is seldom feasible, we consider the effect of numerical resolution on particle clumping and $Z_{\rm crit}$.  We perform our resolution study on the $(\uptau_{\rm s},\textbf{} Z) = (0.01, 0.0133)$ case, which is of special interest.  This $Z$ value lies just below the clumping boundary and $\uptau_{\rm s} = 0.01$ is on the small particle side of the sharp transition in $Z_{\rm crit} (\uptau_{\rm s})$.  We can thus test if these results are resolution dependent.  Furthermore,  Figure \ref{fig:eg_s_max} (left panel) and \ref{fig:s_max} (right panel) show that higher resolution could resolve significantly faster growing linear (unstratified) modes for this case.

We perform runs \texttt{Z1.33t1-2x} and \texttt{Z1.33t1-4x} with double and quadruple the standard resolution, respectively, and in smaller boxes to reduce computational cost (see Table \ref{tab:paras_2D}).  Figure \ref{fig:rhop_xz_resT} shows the saturated clumping state of the three runs of this resolution study.  All cases produce regularly spaced weak clumps, which (as discussed in Section \ref{subsec:boundary}) seems to inhibit stronger clumping.  With increasing resolution, the clump spacing first narrows then widens, while clumping slightly weakens and $H_{\rm p}$ slightly increases (see Table \ref{tab:prop_2D}).  These trends could be due to increased resolution and/or decreased box size, which would require more study and computational resources to disentangle.   

The fact that higher resolution still fails to produce strong clumping nevertheless has interesting implications.  The jump in $Z_{\rm crit}$ for $\uptau_{\rm s} = 0.01$ (vs. $\uptau_{\rm s} = 0.02$) persists despite a factor of 4 increase in resolution.  Since the higher resolution runs should resolve faster growing linear modes, the lack of clumping supports our finding (in Section \ref{subsec:comp_s}) that unstratified linear growth rates fail to predict non-linear clumping in stratified simulations.  

To place our resolution test in context, most previous works show that higher resolution in SI simulations produces stronger clumping.  This trend has been demonstrated for unstratified \citep{Bai2010} and stratified simulations (\citetalias{Yang2017};   our standard resolution matches the recommendation of this work).  Furthermore, \citet{Mignone2019} found faster particle clumping at higher resolution in unstratified, radially global simulations.  Rare cases where higher resolution gives weaker clumping (without the confounding effects of different box sizes) do exist.   \citet{Bai2010a} conducted multi-species, stratified SI simulations with two resolutions.  In only one case (their \texttt{R30Z3-3D} runs) was there strong clumping at only the lower resolution (half our standard resolution).  In \citetalias{Yang2017}, the $\uptau_{\rm s} = 0.01$, $Z = 0.04$ simulations in small boxes ($L_z = L_z = 0.2H$) showed slightly reduced clumping at the highest resolution, though the clumping was increasing sharply at the end of this simulation.  In both cases, the stochastic nature of clumping and finite duration of the simulations may be responsible for these exceptions to the usual increase in SI clumping with resolution.  

Thus, while higher resolution generally favors strong SI clumping, our limited resolution test -- which targets the case where $Z_{\rm crit}$ seems discontinuously large -- suggests that a significant lowering of $Z_{\rm crit}$ with higher resolution is unlikely.  More investigation is certainly warranted.

%%%%%%%%%%%%%%%%%%%%%%%%%%%%%%%%%%%%%%%%%%%%%%%%%%%%%%%%%%%%
\subsubsection{Comparison to 3D Models}
\label{subsubsec:comp_3D}

To test the 2D approximation, we conduct a 3D version of the simulation with $\uptau_{\rm s} = 0.1$ and $Z = 0.75\%$ (Run \texttt{Z0.75t10}).  Ideally, we would perform a parameter survey in 3D as well, but the computational cost for such high resolution simulations is prohibitive.  The case chosen is a good test because it lies near our revised clumping boundary, and probes the interesting regime of $Z < 1\%$ in 3D.

Our 3D run keeps the same grid resolution, with some changes to mitigate computational costs.  The vertical size of the domain was reduced, and a total size $L_X\times L_Y\times L_Z=0.8H\times 0.1H\times 0.15H$ was chosen.  The number of particles is  $2^{26}\approx 6.71\times 10^{7}$, which is larger than the 2D run in total, but a slight reduction in particles per cell (half as many when considering particles per cell within a fixed distance [say $\eta r$] of the midplane).  This particle resolution is consistent with previous 3D work (\citetalias{Li2018}, see their Equation 5). Finally, the run time for the 3D case is reduced to $1000/\Omega_0$.

Figure \ref{fig:parStats_2D3D} shows that both simulations produce strong particle clumping on similar timescales.  It is reassuring that, in 3D simulations, particle clumping still occurs at reduced metallicity compared to previous work. 
The peak particle density and $H_{\rm p}$ show stochastic time variations in both 2D and 3D, a characteristic of larger $\uptau_{\rm s}$ runs.  In the 3D run, average $H_{\rm p}$ is lower and peak $\rho_{\rm p}$ values are somewhat higher and more persistent, compared to 2D.  While the azimuthal direction could introduce non-axisymmetric shearing instabilities that increase stirring, instead extra settling and clumping occur.  For their $\uptau_{\rm s} = 10^{-3}$, $Z = 0.04$ simulations, \citetalias{Yang2017} similarly found stronger clumping in 3D versus 2D runs at the same resolution, $L_X$ and $L_Z$. We thus speculate that a detailed parameter survey of 3D simulations could yield lower $Z_{\rm crit}$ values than reported in this work.

Comparing the 2D and 3D runs, Figure \ref{fig:rhop_xt_2D3D} shows interesting differences in the radial structure of dense filaments (integrated over height and azimuth) versus time.  In 2D, a narrow dense filament sporadically appears and dissolves.  This dense filament interacts with weaker overdensities, and the faster drift (due to weaker feedback) of these weak overdensities is evident in the more vertical orientation of the lower density structures in the plot. In 3D, a different ``filaments-within-filaments'' structure arises.  The larger  ``parent''  contains many higher density sub-filaments within.  While the sub-filaments come and go, the parent filament appears persistent.  Moreover, the void regions outside the parent filament are significantly more particle-depleted in 3D compared to 2D. More study is needed to determine how these 3D structures vary across parameter space, as a great diversity of structures is already seen in 2D (Figure \ref{fig:rhop_xz_tauZ}).

%%%%%%%%%%%%%%%%%%%%%%%%%%%%%%%%%%%%%%%%%%%%%%%%%%%%%%%%%%%%
\subsubsection{Vertical Boundary Conditions and Box Size}
\label{subsubsec:comp_vBCs_Lz}

The importance of box size (\citealp{Yang2014}); \citetalias{Li2018}) and vertical BCs (vBCs; \citetalias{Li2018}) on stratified SI simulations has been analyzed.  Here we reconsider these effects closer to the clumping boundary.

Larger boxes (in all dimensions) are generally considered more realistic (at a given resolution) as box edges introduce artifacts that can inhibit clumping.  In SI simulations, particle-gas interactions very close to the midplane perturb the gas far from the midplane.  With outflow vBCs, these perturbed gas motions are allowed to leave the system.  With periodic (or reflecting)  vBCs, perturbed gas artificially returns to the midplane to provide additional particle stirring.  

To demonstrate the importance of vBCs near the clumping boundary, we consider simulations with $\uptau_{\rm s} = 0.1$ or $0.05$ and $Z = 0.01$ or $ 0.0075$ ($4$ combinations).  For these choices, Figure \ref{fig:comp_ou_pe} compares results using our preferred outflow vBCs to periodic vBCs.  With fixed physical parameters, the periodic cases show evidence of additional particle stirring with larger $H_{\rm p}$, consistent with \citetalias{Li2018}.  For the lower $Z = 0.0075$ runs, the outflow cases give strong clumping while the periodic cases do not produce strong clumping.  This test explicitly demonstrates that outflow vBCs help to lower $Z_{\rm crit}$.

To demonstrate the importance of box height, we compare our standard runs (with $L_Z=0.4H$) to shorter box versions (with $L_Z=0.2H$) for $Z = 0.01$  and $\uptau_{\rm s} = 0.05$, $0.03$, or $0.02$.  Figure \ref{fig:comp_Lz} shows that our standard runs consistently produce lower $H_{\rm p}$.  The effect on clumping is most evident for the $\uptau_{\rm s}=0.02$, where strong clumping in the tall box is faster and stronger.  This test validates our choice of $L_Z=0.4H$ and demonstrates that taller boxes can help to lower $Z_{\rm crit}$.

%%%%%%%%%%%%%%%%%%%%%%%%%%%%%%%%%%%%%%%%%%%%%%%%%%%%%%%%%%%%
\subsection{Comparison with Previous Studies}
\label{subsec:comp_pre}

As noted in Section \ref{subsec:boundary}, our simulations find a clumping threshold, $Z_{\rm crit}$, that is lower than the results of \citetalias{Carrera2015}, but roughly consistent with (and still slightly lower than) the \citetalias{Yang2017} results for $\uptau_{\rm s} = 0.001$ and $0.01$ (see  Figure \ref{fig:map}).  We now compare to these previous works in more detail.  Note that \citetalias{Yang2017} did not revisit the clumping threshold for $\uptau_{\rm s} > 0.01$.  Thus, for $\uptau_{\rm s} > 0.01$ we only need to compare to \citetalias{Carrera2015}, and for $\uptau_{\rm s} \leq 0.01$ it is best to compare to the updated \citetalias{Yang2017} results.

The \citetalias{Carrera2015} study is similar to ours in finding a $Z_{\rm crit}$ value from 2D SI simulations with a range of $\uptau_{\rm s}$ values, and has the same $\Pi$ ($= 0.05$) for their main clumping analysis.   The most significant difference is that \citetalias{Carrera2015} gradually increased $Z$, with a doubling time of 50 orbits, during runs of a given $\uptau_{\rm s}$, while we ran separate simulations with fixed values of $Z$.  When \citetalias{Carrera2015} reported clumping at a given $Z$ value, it means they found significant clumping during a 25 orbit interval (over which $Z$ evolved).  This technique can  overestimate the reported $Z_{\rm crit}$, especially when the clumping timescale is longer than 25 orbits, as we find in many of our fixed-$Z$ simulations (see Figures \ref{fig:eg_H_p_epsilon}, \ref{fig:stiffness}, \ref{fig:parStats_2D3D}, and \ref{fig:comp_ou_pe}).  Compared to simulations with $Z$ increasing in time, our simulations at fixed $Z$ can find clumping at lower $Z$.

\citetalias{Carrera2015} also use a different definition of strong clumping.  They consider not peak densities, but time-averaged deviations from a uniform surface density over the whole box (see their paper for details).  Due to time-averaging, their approach reduces or erases rapidly drifting clumps, which is significant for larger $\uptau_{\rm s}$.  This difference helps explain why our $Z_{\rm crit}$ is so much lower than in \citetalias{Carrera2015} for  $\uptau_{\rm s} \gtrsim 0.3$.  We include clumps that drift, as they could collapse into planetesimals, though 3D self-gravitating simulations are needed to determine which clumps will actually collapse. 

Other differences with \citetalias{Carrera2015} likely contribute to our lower $Z_{\rm crit}$ findings as well.  Our standard grid resolution is twice as high, and our box sizes are all larger than the $L_X = L_Z = 0.2H$ of \citetalias{Carrera2015} in at least one dimension (except for our single $4\times$ resolution test, see Table \ref{tab:paras_2D}).  Our simulations use outflow vertical BCs, compared to the periodic BCs in \citetalias{Carrera2015}.  As described in Section \ref{subsubsec:comp_vBCs_Lz}, increased box size and outflow vBCs naturally promote particle sedimentation and clumping (see Figures \ref{fig:comp_ou_pe} and \ref{fig:comp_Lz}).

\citetalias{Yang2017} assessed the clumping properties of smaller particles ($\uptau_{\rm s} = 0.01$ and $0.001$) using an improved algorithm \citep{Yang2016} to handle stiff drag forces.  While our clumping results are similar to \citetalias{Yang2017}, we did find slightly lower $Z_{\rm crit}$ (see Figure \ref{fig:map}) and it is worth summarizing the main similarities and differences.   \citetalias{Yang2017} also ran simulations with fixed $Z$ values and $\Pi = 0.05$.  Their analysis also used $\rho_{\rm p, max}$ to quantify strong clumping, but \citetalias{Yang2017} adopted a more generous threshold, with $\rho_{\rm p, max} \gtrsim 30 \rho_{\rm g}$ considered strong clumping (their $\uptau_{\rm s} = 0.01$, $Z = 0.02$ case).   \citetalias{Yang2017} used periodic vBCs, which is probably the largest reason their $Z_{\rm crit}$ are slightly larger than our results, as discussed above.

In their suite of 2D simulations, \citetalias{Yang2017} made different trade-offs in resolution vs.\ box size.  \citetalias{Yang2017} performed a larger number of high resolution simulations, at $2560/H$ and $5120/H$ [i.e.\ $128/(\eta r)$ and $256/(\eta r)$], equal to our resolution test.  However, their higher resolution runs were in smaller $L_X = L_Z = 0.2H$ boxes (except for a single $\uptau_{\rm s} = 0.01$, $Z = 0.02$ run at $2560/H$ ).  Both works considered long simulation times to allow clumps to develop. 

One interesting difference with \citetalias{Yang2017} is the qualitative appearance of the midplane layers.  For  $\uptau_{\rm s} = 10^{-3}$, their Figure 4 does not show the ``fish-like'' structures seen in our Figure \ref{fig:rhop_xz_tauZ} (top panel, with more cases at \href{https://rixinli.com/tauZmap.html}{https://rixinli.com/tauZmap.html}).  For  $\uptau_{\rm s} = 10^{-2}$, the differences are more subtle, but noticeable away from the midplane at lower dust densities.  Figure 1 of \citetalias{Yang2017}  shows vertically-extended structures of low dust density that can reach their vertical boundaries.  By comparison in our Figure \ref{fig:rhop_xz_resT}) (also \ref{fig:rhop_xz_tauZ}, second panel from the top)  the corresponding low density structures have a more disordered appearance and do not extend as far from the midplane.  Indeed, our $H_{\rm p}$ values are generally lower than \citetalias{Yang2017} found.  The differences in vBCs likely play a role in these structural differences, but we cannot rule out other code or algorithm differences at this point. 

Both \citetalias{Carrera2015} and \citetalias{Yang2017} used (different versions of) the \texttt{Pencil} code, while we use \texttt{ATHENA}.  A detailed code comparison test (which is not our goal) would be needed to attribute any differences to the algorithms, and we note that both codes have been extensively tested.  Nevertheless, this study  motivates specific tests of non-linear SI convergence for a range of codes.

Finally, we note that both \citetalias{Carrera2015} and \citetalias{Yang2017} used $n_p = 1$ particle per cell, compared to our choice of $n_p = 4$.  As described in Section \ref{sec:method}, we probably used more particles than needed, so their $n_p = 1$ choice should be sufficient.

%%%%%%%%%%%%%%%%%%%%%%%%%%%%%%%%%%%%%%%%%%%%%%%%%%%%%%%%%%%%%%%%%%%%%%%%%%%%%%%
\section{Conclusions}
\label{sec:conclusions}

We study the thresholds for strong particle clumping by the streaming instability with a suite of high resolution simulations in stratified disk models.  This work neglects  self-gravity to focus on particle concentration by the streaming instability.  Particle concentration occurs above a critical particle-to-gas surface density ratio, $Z_{\rm crit}$, which we characterize as a function of particle stopping time $\uptau_{\rm s}$ (a proxy for size).  Our main results are as follows:

\begin{enumerate}
  \item We find lower $Z_{\rm crit}$ for particle concentration by the SI than previous works, especially for intermediate-sized solids ($\uptau_{\rm s} > 0.01$).  In smooth disks with a standard pressure gradient parameter ($\Pi = 0.05$), strong particle clumping occurs at sub-solar metallicities ($Z_{\rm crit} < 0.01$) for $\uptau_{\rm s}  > 0.01$, with the smallest $Z_{\rm crit}  = 0.4\%$ for solids with $\uptau_{\rm s} = 0.3$.  Smaller particles, with $\uptau_{\rm s} \leq 0.01$ require super-solar $Z_{\rm crit} \gtrsim 0.02$.  Our revised clumping boundary is shown in Figure \ref{fig:map}.
  \item We fit our updated $Z_{\rm crit}$ and generalize to a range of pressure gradients in  Equation \ref{eq:Z_crit}.
  \item We further generalize to include the effects of additional (non-SI) turbulence as $Z_{\rm crit, \alpha}$ in Equation \ref{eq:Z_alpha}.  Our $Z_{\rm crit, \alpha}$ improves on previous estimates of SI clumping threshold with turbulence.  We use our numerical simulations to determine the required midplane dust-to-gas ratios as a function of size, $\epsilon_{\rm crit}(\uptau_{\rm s})$.  Previous works assume $\epsilon_{\rm crit} = 1$ for all sizes. We find $\epsilon_{\rm crit} \lesssim 0.5$ for intermediate sizes ($0.02 < \uptau_{\rm s} < 1$) and  $\epsilon_{\rm crit} \sim 2.5$ for  $\uptau_{\rm s} \leq 0.01$, as shown in Figure \ref{fig:H_p_epsilon} and Equation \ref{eq:ep_crit}.
  \item A sharp transition in $Z_{\rm crit}$ and in $\epsilon_{\rm crit}$ at $\uptau_{\rm s} = 0.01$ is discovered, with significantly larger values for smaller solids. While particle diffusion, linear growth rates and numerical resolution are investigated as possible causes of the sharp transition, no fundamental explanation is found.
  \item We find that linear, unstratified SI growth rates are not a good predictor for SI clumping in stratified simulations.  Calculated growth rates are surprisingly similar in clumping and non-clumping runs (see Figures \ref{fig:eg_s_max} and \ref{fig:s_max}).  In fact, when turbulent diffusion is included at the level measured in our simulations (see Equation \ref{eq:s_damped}) standard linear theory predicts no SI growth in any of our runs!  We thus urge significant caution when predicting clumping based on linear theory.
\end{enumerate}

Building on previous work on SI clumping thresholds, our results can be used to refine predictions of when protoplanetary disk models -- including models based on disk observations -- should form planetesimals by the SI mechanism.  Future work can make these predictions even more realistic.  Self-gravitating, 3D simulations -- especially near the clumping boundary -- are needed to confirm when strong clumping will actually produce gravitational collapse.   

Furthermore, dust particles in disks are expected to have a size distribution.  The role of size distributions on linear, unstratified SI has already been investigated \citep{Krapp2019, Paardekooper2020, Zhu2021}.  More work is needed on the role of size distributions on particle clumping with particle settling (e.g., Shaffer et al.\ in prep.).

The efficiency of dust growth and resulting size distributions remain uncertain, especially for the young, embedded protoplanetary disks.  However, observations of these Class I sources find evidence for grain growth to mm-sizes \citep{Harsono2018} and for disk substructures that might be caused by already-formed planets \citep{Segura-Cox2020, Cieza2021}.  Our results for SI-induced particle concentration at low metallicity are consistent with these observations, and with a scenario of early planet formation in the core accretion hypothesis \citep{Youdin2013}. 

%%%%%%%%%%%%%%%%%%%%%%%%%%%%%%%%%%%%%%%%%%%%%%%%%%%%%%%%%%%%%%%%%%%%%%%%%%%%%%%%
\section*{Acknowledgements}

We thank the anonymous referee for useful suggestions.  We thank Chao-chin Yang, Anders Johansen, Zhaohuan Zhu, Kaitlin Kratter, and Leonardo Krapp for useful discussions.  RL acknowledges support from NASA headquarters under the NASA Earth and Space Science Fellowship Program grant NNX16AP53H. ANY acknowledges support from NASA through awards NNX17AK59G and 80NSSC21K0497 and from the NSF through grant AST-1616929.

\added{\software{ATHENA \citep{Stone2008, Bai2010}, 
          Matplotlib \citep{Matplotlib}, 
          Numpy \& Scipy \citep{NumPy, SciPy},
          Pyridoxine \citep{Pyridoxine},
          PLAN \citep{PLAN}.}}

%%%%%%%%%%%%%%%%%%%% REFERENCES %%%%%%%%%%%%%%%%%%
\bibliographystyle{aasjournal}
\bibliography{refs}

\begin{thebibliography}{}
\expandafter\ifx\csname natexlab\endcsname\relax\def\natexlab#1{#1}\fi
\providecommand{\url}[1]{\href{#1}{#1}}
\providecommand{\dodoi}[1]{doi:~\href{http://doi.org/#1}{\nolinkurl{#1}}}
\providecommand{\doeprint}[1]{\href{http://ascl.net/#1}{\nolinkurl{http://ascl.net/#1}}}
\providecommand{\doarXiv}[1]{\href{https://arxiv.org/abs/#1}{\nolinkurl{https://arxiv.org/abs/#1}}}

\bibitem[{{Adachi} {et~al.}(1976){Adachi}, {Hayashi}, \&
  {Nakazawa}}]{Adachi1976}
{Adachi}, I., {Hayashi}, C., \& {Nakazawa}, K. 1976, Progress of Theoretical
  Physics, 56, 1756, \dodoi{10.1143/PTP.56.1756}

\bibitem[{{ALMA Partnership} {et~al.}(2015){ALMA Partnership}, {Brogan},
  {P{\'e}rez}, {Hunter}, {Dent}, {Hales}, {Hills}, {Corder}, {Fomalont},
  {Vlahakis}, {Asaki}, {Barkats}, {Hirota}, {Hodge}, {Impellizzeri}, {Kneissl},
  {Liuzzo}, {Lucas}, {Marcelino}, {Matsushita}, {Nakanishi}, {Phillips},
  {Richards}, {Toledo}, {Aladro}, {Broguiere}, {Cortes}, {Cortes}, {Espada},
  {Galarza}, {Garcia-Appadoo}, {Guzman-Ramirez}, {Humphreys}, {Jung}, {Kameno},
  {Laing}, {Leon}, {Marconi}, {Mignano}, {Nikolic}, {Nyman}, {Radiszcz},
  {Remijan}, {Rod{\'o}n}, {Sawada}, {Takahashi}, {Tilanus}, {Vila Vilaro},
  {Watson}, {Wiklind}, {Akiyama}, {Chapillon}, {de Gregorio-Monsalvo}, {Di
  Francesco}, {Gueth}, {Kawamura}, {Lee}, {Nguyen Luong}, {Mangum}, {Pietu},
  {Sanhueza}, {Saigo}, {Takakuwa}, {Ubach}, {van Kempen}, {Wootten},
  {Castro-Carrizo}, {Francke}, {Gallardo}, {Garcia}, {Gonzalez}, {Hill},
  {Kaminski}, {Kurono}, {Liu}, {Lopez}, {Morales}, {Plarre}, {Schieven},
  {Testi}, {Videla}, {Villard}, {Andreani}, {Hibbard}, \&
  {Tatematsu}}]{ALMAPartnership2015}
{ALMA Partnership}, {Brogan}, C.~L., {P{\'e}rez}, L.~M., {et~al.} 2015, \apjl,
  808, L3, \dodoi{10.1088/2041-8205/808/1/L3}

\bibitem[{{Andrews}(2020)}]{Andrews2020}
{Andrews}, S.~M. 2020, \araa, 58, 483,
  \dodoi{10.1146/annurev-astro-031220-010302}

\bibitem[{{Andrews} {et~al.}(2016){Andrews}, {Wilner}, {Zhu}, {Birnstiel},
  {Carpenter}, {P{\'e}rez}, {Bai}, {{\"O}berg}, {Hughes}, {Isella}, \&
  {Ricci}}]{Andrews2016}
{Andrews}, S.~M., {Wilner}, D.~J., {Zhu}, Z., {et~al.} 2016, \apjl, 820, L40,
  \dodoi{10.3847/2041-8205/820/2/L40}

\bibitem[{{Andrews} {et~al.}(2018){Andrews}, {Huang}, {P{\'e}rez}, {Isella},
  {Dullemond}, {Kurtovic}, {Guzm{\'a}n}, {Carpenter}, {Wilner}, {Zhang}, {Zhu},
  {Birnstiel}, {Bai}, {Benisty}, {Hughes}, {{\"O}berg}, \&
  {Ricci}}]{Andrews2018}
{Andrews}, S.~M., {Huang}, J., {P{\'e}rez}, L.~M., {et~al.} 2018, ApJL, 869,
  L41, \dodoi{10.3847/2041-8213/aaf741}

\bibitem[{{Asplund} {et~al.}(2009){Asplund}, {Grevesse}, {Sauval}, \&
  {Scott}}]{2009Asplund}
{Asplund}, M., {Grevesse}, N., {Sauval}, A.~J., \& {Scott}, P. 2009, \araa, 47,
  481, \dodoi{10.1146/annurev.astro.46.060407.145222}

\bibitem[{{Bai} \& {Stone}(2010{\natexlab{a}})}]{Bai2010}
{Bai}, X.-N., \& {Stone}, J.~M. 2010{\natexlab{a}}, ApJS, 190, 297,
  \dodoi{10.1088/0067-0049/190/2/297}

\bibitem[{{Bai} \& {Stone}(2010{\natexlab{b}})}]{Bai2010a}
---. 2010{\natexlab{b}}, ApJ, 722, 1437, \dodoi{10.1088/0004-637X/722/2/1437}

\bibitem[{{Blum}(2018)}]{Blum2018}
{Blum}, J. 2018, SSR, 214, 52, \dodoi{10.1007/s11214-018-0486-5}

\bibitem[{{Blum} {et~al.}(2017){Blum}, {Gundlach}, {Krause}, {Fulle},
  {Johansen}, {Agarwal}, {von Borstel}, {Shi}, {Hu}, {Bentley}, {Capaccioni},
  {Colangeli}, {Della Corte}, {Fougere}, {Green}, {Ivanovski}, {Mannel},
  {Merouane}, {Migliorini}, {Rotundi}, {Schmied}, \& {Snodgrass}}]{Blum2017}
{Blum}, J., {Gundlach}, B., {Krause}, M., {et~al.} 2017, \mnras, 469, S755,
  \dodoi{10.1093/mnras/stx2741}

\bibitem[{{Carrera} {et~al.}(2015){Carrera}, {Johansen}, \&
  {Davies}}]{Carrera2015}
{Carrera}, D., {Johansen}, A., \& {Davies}, M.~B. 2015, A\&A, 579, A43,
  \dodoi{10.1051/0004-6361/201425120}

\bibitem[{{Carrera} {et~al.}(2021){Carrera}, {Simon}, {Li}, {Kretke}, \&
  {Klahr}}]{Carrera2021}
{Carrera}, D., {Simon}, J.~B., {Li}, R., {Kretke}, K.~A., \& {Klahr}, H. 2021,
  \aj, 161, 96, \dodoi{10.3847/1538-3881/abd4d9}

\bibitem[{{Chiang} \& {Youdin}(2010)}]{Chiang2010}
{Chiang}, E., \& {Youdin}, A.~N. 2010, Annual Review of Earth and Planetary
  Sciences, 38, 493, \dodoi{10.1146/annurev-earth-040809-152513}

\bibitem[{{Cieza} {et~al.}(2021){Cieza}, {Gonz{\'a}lez-Ruilova}, {Hales},
  {Pinilla}, {Ru{\'\i}z-Rodr{\'\i}guez}, {Zurlo}, {Casassus}, {P{\'e}rez},
  {C{\'a}novas}, {Arce-Tord}, {Flock}, {Kurtovic}, {Marino}, {Nogueira},
  {Perez}, {Price}, {Principe}, \& {Williams}}]{Cieza2021}
{Cieza}, L.~A., {Gonz{\'a}lez-Ruilova}, C., {Hales}, A.~S., {et~al.} 2021,
  \mnras, 501, 2934, \dodoi{10.1093/mnras/staa3787}

\bibitem[{{Dong} {et~al.}(2015){Dong}, {Zhu}, \& {Whitney}}]{Dong2015}
{Dong}, R., {Zhu}, Z., \& {Whitney}, B. 2015, \apj, 809, 93,
  \dodoi{10.1088/0004-637X/809/1/93}

\bibitem[{{Dr{\k{a}}{\.z}kowska} {et~al.}(2016){Dr{\k{a}}{\.z}kowska},
  {Alibert}, \& {Moore}}]{Drazkowska2016}
{Dr{\k{a}}{\.z}kowska}, J., {Alibert}, Y., \& {Moore}, B. 2016, A\&A, 594,
  A105, \dodoi{10.1051/0004-6361/201628983}

\bibitem[{{Dr{\k{a}}{\.z}kowska} \& {Dullemond}(2014)}]{Drazkowska2014}
{Dr{\k{a}}{\.z}kowska}, J., \& {Dullemond}, C.~P. 2014, A\&A, 572, A78,
  \dodoi{10.1051/0004-6361/201424809}

\bibitem[{{Dubrulle} {et~al.}(1995){Dubrulle}, {Morfill}, \&
  {Sterzik}}]{Dubrulle1995}
{Dubrulle}, B., {Morfill}, G., \& {Sterzik}, M. 1995, \icarus, 114, 237,
  \dodoi{10.1006/icar.1995.1058}

\bibitem[{{Dullemond} {et~al.}(2018){Dullemond}, {Birnstiel}, {Huang},
  {Kurtovic}, {Andrews}, {Guzm{\'a}n}, {P{\'e}rez}, {Isella}, {Zhu}, {Benisty},
  {Wilner}, {Bai}, {Carpenter}, {Zhang}, \& {Ricci}}]{Dullemond2018}
{Dullemond}, C.~P., {Birnstiel}, T., {Huang}, J., {et~al.} 2018, ApJL, 869,
  L46, \dodoi{10.3847/2041-8213/aaf742}

\bibitem[{{Gole} {et~al.}(2020){Gole}, {Simon}, {Li}, {Youdin}, \&
  {Armitage}}]{Gole2020}
{Gole}, D.~A., {Simon}, J.~B., {Li}, R., {Youdin}, A.~N., \& {Armitage}, P.~J.
  2020, \apj, 904, 132, \dodoi{10.3847/1538-4357/abc334}

\bibitem[{{Goodman} \& {Pindor}(2000)}]{Goodman2000}
{Goodman}, J., \& {Pindor}, B. 2000, Icarus, 148, 537,
  \dodoi{10.1006/icar.2000.6467}

\bibitem[{Harris {et~al.}(2020)Harris, Millman, van~der Walt, Gommers,
  Virtanen, Cournapeau, Wieser, Taylor, Berg, Smith, Kern, Picus, Hoyer, van
  Kerkwijk, Brett, Haldane, del R{\'{i}}o, Wiebe, Peterson,
  G{\'{e}}rard-Marchant, Sheppard, Reddy, Weckesser, Abbasi, Gohlke, \&
  Oliphant}]{NumPy}
Harris, C.~R., Millman, K.~J., van~der Walt, S.~J., {et~al.} 2020, Nature, 585,
  357, \dodoi{10.1038/s41586-020-2649-2}

\bibitem[{{Harsono} {et~al.}(2018){Harsono}, {Bjerkeli}, {van der Wiel},
  {Ramsey}, {Maud}, {Kristensen}, \& {J{\o}rgensen}}]{Harsono2018}
{Harsono}, D., {Bjerkeli}, P., {van der Wiel}, M. H.~D., {et~al.} 2018, Nature
  Astronomy, 2, 646, \dodoi{10.1038/s41550-018-0497-x}

\bibitem[{{Hawley} {et~al.}(1995){Hawley}, {Gammie}, \& {Balbus}}]{Hawley1995}
{Hawley}, J.~F., {Gammie}, C.~F., \& {Balbus}, S.~A. 1995, ApJ, 440, 742,
  \dodoi{10.1086/175311}

\bibitem[{Hunter(2007)}]{Matplotlib}
Hunter, J.~D. 2007, Computing in Science Engineering, 9, 90,
  \dodoi{10.1109/MCSE.2007.55}

\bibitem[{{Jin} {et~al.}(2016){Jin}, {Li}, {Isella}, {Li}, \& {Ji}}]{Jin2016}
{Jin}, S., {Li}, S., {Isella}, A., {Li}, H., \& {Ji}, J. 2016, \apj, 818, 76,
  \dodoi{10.3847/0004-637X/818/1/76}

\bibitem[{{Johansen} {et~al.}(2014){Johansen}, {Blum}, {Tanaka}, {Ormel},
  {Bizzarro}, \& {Rickman}}]{Johansen2014}
{Johansen}, A., {Blum}, J., {Tanaka}, H., {et~al.} 2014, Protostars and Planets
  VI, 547, \dodoi{10.2458/azu_uapress_9780816531240-ch024}

\bibitem[{{Johansen} {et~al.}(2015){Johansen}, {Mac Low}, {Lacerda}, \&
  {Bizzarro}}]{Johansen2015}
{Johansen}, A., {Mac Low}, M.-M., {Lacerda}, P., \& {Bizzarro}, M. 2015,
  Science Advances, 1, 1500109, \dodoi{10.1126/sciadv.1500109}

\bibitem[{{Johansen} {et~al.}(2007){Johansen}, {Oishi}, {Mac Low}, {Klahr},
  {Henning}, \& {Youdin}}]{Johansen2007a}
{Johansen}, A., {Oishi}, J.~S., {Mac Low}, M.-M., {et~al.} 2007, Nature, 448,
  1022, \dodoi{10.1038/nature06086}

\bibitem[{{Johansen} \& {Youdin}(2007)}]{Johansen2007}
{Johansen}, A., \& {Youdin}, A. 2007, \apj, 662, 627, \dodoi{10.1086/516730}

\bibitem[{{Johansen} {et~al.}(2009{\natexlab{a}}){Johansen}, {Youdin}, \&
  {Klahr}}]{Johansen2009}
{Johansen}, A., {Youdin}, A., \& {Klahr}, H. 2009{\natexlab{a}}, ApJ, 697,
  1269, \dodoi{10.1088/0004-637X/697/2/1269}

\bibitem[{{Johansen} {et~al.}(2009{\natexlab{b}}){Johansen}, {Youdin}, \& {Mac
  Low}}]{Johansen2009a}
{Johansen}, A., {Youdin}, A., \& {Mac Low}, M.-M. 2009{\natexlab{b}}, ApJL,
  704, L75, \dodoi{10.1088/0004-637X/704/2/L75}

\bibitem[{{Johansen} {et~al.}(2012){Johansen}, {Youdin}, \&
  {Lithwick}}]{Johansen2012}
{Johansen}, A., {Youdin}, A.~N., \& {Lithwick}, Y. 2012, A\&A, 537, A125,
  \dodoi{10.1051/0004-6361/201117701}

\bibitem[{{Krapp} {et~al.}(2019){Krapp}, {Ben{\'\i}tez-Llambay}, {Gressel}, \&
  {Pessah}}]{Krapp2019}
{Krapp}, L., {Ben{\'\i}tez-Llambay}, P., {Gressel}, O., \& {Pessah}, M.~E.
  2019, ApJL, 878, L30, \dodoi{10.3847/2041-8213/ab2596}

\bibitem[{{Krapp} {et~al.}(2020){Krapp}, {Youdin}, {Kratter}, \&
  {Ben{\'\i}tez-Llambay}}]{Krapp2020}
{Krapp}, L., {Youdin}, A.~N., {Kratter}, K.~M., \& {Ben{\'\i}tez-Llambay}, P.
  2020, \mnras, 497, 2715, \dodoi{10.1093/mnras/staa1854}

\bibitem[{{Li}(2019)}]{PLAN}
{Li}, R. 2019, {PLAN: A Clump-finder for Planetesimal Formation Simulations}.
\newblock \doeprint{1911.001}

\bibitem[{Li(2020)}]{Pyridoxine}
Li, R. 2020, Pyridoxine: Handy Python Snippets for Athena Data.
\newblock \url{https://pypi.org/project/pyridoxine}

\bibitem[{{Li} {et~al.}(2018){Li}, {Youdin}, \& {Simon}}]{Li2018}
{Li}, R., {Youdin}, A.~N., \& {Simon}, J.~B. 2018, ApJ, 862, 14,
  \dodoi{10.3847/1538-4357/aaca99}

\bibitem[{{Li} {et~al.}(2019){Li}, {Youdin}, \& {Simon}}]{Li2019}
---. 2019, \apj, 885, 69, \dodoi{10.3847/1538-4357/ab480d}

\bibitem[{{Lin}(2021)}]{Lin2021}
{Lin}, M.-K. 2021, \apj, 907, 64, \dodoi{10.3847/1538-4357/abcd9b}

\bibitem[{{Lin} \& {Youdin}(2017)}]{Lin2017}
{Lin}, M.-K., \& {Youdin}, A.~N. 2017, ApJ, 849, 129,
  \dodoi{10.3847/1538-4357/aa92cd}

\bibitem[{{Liu} {et~al.}(2019){Liu}, {Ormel}, \& {Johansen}}]{Liu2019}
{Liu}, B., {Ormel}, C.~W., \& {Johansen}, A. 2019, \aap, 624, A114,
  \dodoi{10.1051/0004-6361/201834174}

\bibitem[{{Long} {et~al.}(2018){Long}, {Pinilla}, {Herczeg}, {Harsono},
  {Dipierro}, {Pascucci}, {Hendler}, {Tazzari}, {Ragusa}, {Salyk}, {Edwards},
  {Lodato}, {van de Plas}, {Johnstone}, {Liu}, {Boehler}, {Cabrit}, {Manara},
  {Menard}, {Mulders}, {Nisini}, {Fischer}, {Rigliaco}, {Banzatti}, {Avenhaus},
  \& {Gully-Santiago}}]{Long2018}
{Long}, F., {Pinilla}, P., {Herczeg}, G.~J., {et~al.} 2018, \apj, 869, 17,
  \dodoi{10.3847/1538-4357/aae8e1}

\bibitem[{{Lovelace} {et~al.}(1999){Lovelace}, {Li}, {Colgate}, \&
  {Nelson}}]{Lovelace1999}
{Lovelace}, R.~V.~E., {Li}, H., {Colgate}, S.~A., \& {Nelson}, A.~F. 1999,
  \apj, 513, 805, \dodoi{10.1086/306900}

\bibitem[{{McKinnon} {et~al.}(2020){McKinnon}, {Richardson}, {Marohnic},
  {Keane}, {Grundy}, {Hamilton}, {Nesvorn{\'y}}, {Umurhan}, {Lauer}, {Singer},
  {Stern}, {Weaver}, {Spencer}, {Buie}, {Moore}, {Kavelaars}, {Lisse}, {Mao},
  {Parker}, {Porter}, {Showalter}, {Olkin}, {Cruikshank}, {Elliott},
  {Gladstone}, {Parker}, {Verbiscer}, {Young}, \& {New Horizons Science
  Team}}]{McKinnon2020}
{McKinnon}, W.~B., {Richardson}, D.~C., {Marohnic}, J.~C., {et~al.} 2020,
  Science, 367, aay6620, \dodoi{10.1126/science.aay6620}

\bibitem[{Mignone {et~al.}(2019)Mignone, Flock, \& Vaidya}]{Mignone2019}
Mignone, A., Flock, M., \& Vaidya, B. 2019, The Astrophysical Journal
  Supplement Series, 244, 38, \dodoi{10.3847/1538-4365/ab4356}

\bibitem[{{Nakagawa} {et~al.}(1986){Nakagawa}, {Sekiya}, \&
  {Hayashi}}]{Nakagawa1986}
{Nakagawa}, Y., {Sekiya}, M., \& {Hayashi}, C. 1986, Icarus, 67, 375,
  \dodoi{10.1016/0019-1035(86)90121-1}

\bibitem[{{Nesvorn{\'y}} {et~al.}(2021){Nesvorn{\'y}}, {Li}, {Simon}, {Youdin},
  {Richardson}, {Marschall}, \& {Grundy}}]{Nesvorny2021}
{Nesvorn{\'y}}, D., {Li}, R., {Simon}, J.~B., {et~al.} 2021, The Planetary
  Science Journal, 2, 27, \dodoi{10.3847/PSJ/abd858}

\bibitem[{{Nesvorn{\'y}} {et~al.}(2019){Nesvorn{\'y}}, {Li}, {Youdin}, {Simon},
  \& {Grundy}}]{Nesvorny2019}
{Nesvorn{\'y}}, D., {Li}, R., {Youdin}, A.~N., {Simon}, J.~B., \& {Grundy},
  W.~M. 2019, Nature Astronomy, 3, 808, \dodoi{10.1038/s41550-019-0806-z}

\bibitem[{{Nesvorn{\'y}} {et~al.}(2010){Nesvorn{\'y}}, {Youdin}, \&
  {Richardson}}]{Nesvorny2010}
{Nesvorn{\'y}}, D., {Youdin}, A.~N., \& {Richardson}, D.~C. 2010, AJ, 140, 785,
  \dodoi{10.1088/0004-6256/140/3/785}

\bibitem[{{Onishi} \& {Sekiya}(2017)}]{Onishi2017}
{Onishi}, I.~K., \& {Sekiya}, M. 2017, Earth, Planets, and Space, 69, 50,
  \dodoi{10.1186/s40623-017-0637-z}

\bibitem[{{Paardekooper} {et~al.}(2020){Paardekooper}, {McNally}, \&
  {Lovascio}}]{Paardekooper2020}
{Paardekooper}, S.-J., {McNally}, C.~P., \& {Lovascio}, F. 2020, \mnras, 499,
  4223, \dodoi{10.1093/mnras/staa3162}

\bibitem[{{Pinilla} \& {Youdin}(2017)}]{Pinilla2017}
{Pinilla}, P., \& {Youdin}, A. 2017, in Astrophysics and Space Science Library,
  Vol. 445, Astrophysics and Space Science Library, ed. M.~{Pessah} \&
  O.~{Gressel} (Springer International Publishing), 91,
  \dodoi{10.1007/978-3-319-60609-5_4}

\bibitem[{{Pinte} {et~al.}(2018){Pinte}, {Price}, {M{\'e}nard}, {Duch{\^e}ne},
  {Dent}, {Hill}, {de Gregorio-Monsalvo}, {Hales}, \& {Mentiplay}}]{Pinte2018}
{Pinte}, C., {Price}, D.~J., {M{\'e}nard}, F., {et~al.} 2018, \apjl, 860, L13,
  \dodoi{10.3847/2041-8213/aac6dc}

\bibitem[{{Sch{\"a}fer} {et~al.}(2017){Sch{\"a}fer}, {Yang}, \&
  {Johansen}}]{Schafer2017}
{Sch{\"a}fer}, U., {Yang}, C.-C., \& {Johansen}, A. 2017, A\&A, 597, A69,
  \dodoi{10.1051/0004-6361/201629561}

\bibitem[{{Schoonenberg} \& {Ormel}(2017)}]{Schoonenberg2017}
{Schoonenberg}, D., \& {Ormel}, C.~W. 2017, A\&A, 602, A21,
  \dodoi{10.1051/0004-6361/201630013}

\bibitem[{{Segura-Cox} {et~al.}(2020){Segura-Cox}, {Schmiedeke}, {Pineda},
  {Stephens}, {Fern{\'a}ndez-L{\'o}pez}, {Looney}, {Caselli}, {Li}, {Mundy},
  {Kwon}, \& {Harris}}]{Segura-Cox2020}
{Segura-Cox}, D.~M., {Schmiedeke}, A., {Pineda}, J.~E., {et~al.} 2020, \nat,
  586, 228, \dodoi{10.1038/s41586-020-2779-6}

\bibitem[{{Sekiya} \& {Onishi}(2018)}]{Sekiya2018}
{Sekiya}, M., \& {Onishi}, I.~K. 2018, \apj, 860, 140,
  \dodoi{10.3847/1538-4357/aac4a7}

\bibitem[{{Sheehan} {et~al.}(2020){Sheehan}, {Tobin}, {Federman}, {Megeath}, \&
  {Looney}}]{Sheehan2020}
{Sheehan}, P.~D., {Tobin}, J.~J., {Federman}, S., {Megeath}, S.~T., \&
  {Looney}, L.~W. 2020, \apj, 902, 141, \dodoi{10.3847/1538-4357/abbad5}

\bibitem[{{Simon} {et~al.}(2016){Simon}, {Armitage}, {Li}, \&
  {Youdin}}]{Simon2016}
{Simon}, J.~B., {Armitage}, P.~J., {Li}, R., \& {Youdin}, A.~N. 2016, ApJ, 822,
  55, \dodoi{10.3847/0004-637X/822/1/55}

\bibitem[{{Squire} \& {Hopkins}(2018)}]{Squire2018}
{Squire}, J., \& {Hopkins}, P.~F. 2018, \mnras, 477, 5011,
  \dodoi{10.1093/mnras/sty854}

\bibitem[{{Stammler} {et~al.}(2019){Stammler}, {Dr{\k{a}}{\.z}kowska},
  {Birnstiel}, {Klahr}, {Dullemond}, \& {Andrews}}]{Stammler2019}
{Stammler}, S.~M., {Dr{\k{a}}{\.z}kowska}, J., {Birnstiel}, T., {et~al.} 2019,
  \apjl, 884, L5, \dodoi{10.3847/2041-8213/ab4423}

\bibitem[{{Stone} {et~al.}(2008){Stone}, {Gardiner}, {Teuben}, {Hawley}, \&
  {Simon}}]{Stone2008}
{Stone}, J.~M., {Gardiner}, T.~A., {Teuben}, P., {Hawley}, J.~F., \& {Simon},
  J.~B. 2008, ApJS, 178, 137, \dodoi{10.1086/588755}

\bibitem[{{Takeuchi} {et~al.}(2012){Takeuchi}, {Muto}, {Okuzumi}, {Ishitsu}, \&
  {Ida}}]{Takeuchi2012}
{Takeuchi}, T., {Muto}, T., {Okuzumi}, S., {Ishitsu}, N., \& {Ida}, S. 2012,
  \apj, 744, 101, \dodoi{10.1088/0004-637X/744/2/101}

\bibitem[{{Teague} {et~al.}(2019){Teague}, {Bae}, \& {Bergin}}]{Teague2019}
{Teague}, R., {Bae}, J., \& {Bergin}, E.~A. 2019, \nat, 574, 378,
  \dodoi{10.1038/s41586-019-1642-0}

\bibitem[{Virtanen {et~al.}(2020)Virtanen, Gommers, Oliphant, Haberland, Reddy,
  Cournapeau, Burovski, Peterson, Weckesser, Bright, {van der Walt}, Brett,
  Wilson, Millman, Mayorov, Nelson, Jones, Kern, Larson, Carey, Polat, Feng,
  Moore, {VanderPlas}, Laxalde, Perktold, Cimrman, Henriksen, Quintero, Harris,
  Archibald, Ribeiro, Pedregosa, {van Mulbregt}, \& {SciPy 1.0
  Contributors}}]{SciPy}
Virtanen, P., Gommers, R., Oliphant, T.~E., {et~al.} 2020, Nature Methods, 17,
  261, \dodoi{10.1038/s41592-019-0686-2}

\bibitem[{{Yang} \& {Johansen}(2014)}]{Yang2014}
{Yang}, C.-C., \& {Johansen}, A. 2014, ApJ, 792, 86,
  \dodoi{10.1088/0004-637X/792/2/86}

\bibitem[{{Yang} \& {Johansen}(2016)}]{Yang2016}
---. 2016, \apjs, 224, 39, \dodoi{10.3847/0067-0049/224/2/39}

\bibitem[{{Yang} {et~al.}(2017){Yang}, {Johansen}, \& {Carrera}}]{Yang2017}
{Yang}, C.-C., {Johansen}, A., \& {Carrera}, D. 2017, A\&A, 606, A80,
  \dodoi{10.1051/0004-6361/201630106}

\bibitem[{{Youdin} \& {Johansen}(2007)}]{YJ2007}
{Youdin}, A., \& {Johansen}, A. 2007, ApJ, 662, 613, \dodoi{10.1086/516729}

\bibitem[{{Youdin}(2010)}]{Youdin2010}
{Youdin}, A.~N. 2010, in EAS Publications Series, Vol.~41, EAS Publications
  Series, ed. T.~{Montmerle}, D.~{Ehrenreich}, \& A.-M. {Lagrange} (EDP
  Sciences), 187--207, \dodoi{10.1051/eas/1041016}

\bibitem[{{Youdin} \& {Goodman}(2005)}]{Youdin2005}
{Youdin}, A.~N., \& {Goodman}, J. 2005, ApJ, 620, 459, \dodoi{10.1086/426895}

\bibitem[{{Youdin} \& {Kenyon}(2013)}]{Youdin2013}
{Youdin}, A.~N., \& {Kenyon}, S.~J. 2013, {From Disks to Planets}, ed. T.~D.
  {Oswalt}, L.~M. {French}, \& P.~{Kalas} (Dordrecht: Springer Netherlands), 1,
  \dodoi{10.1007/978-94-007-5606-9_1}

\bibitem[{{Youdin} \& {Lithwick}(2007)}]{Youdin2007}
{Youdin}, A.~N., \& {Lithwick}, Y. 2007, \icarus, 192, 588,
  \dodoi{10.1016/j.icarus.2007.07.012}

\bibitem[{{Youdin} \& {Shu}(2002)}]{Youdin2002}
{Youdin}, A.~N., \& {Shu}, F.~H. 2002, ApJ, 580, 494, \dodoi{10.1086/343109}

\bibitem[{{Zhang} {et~al.}(2018){Zhang}, {Zhu}, {Huang}, {Guzm{\'a}n},
  {Andrews}, {Birnstiel}, {Dullemond}, {Carpenter}, {Isella}, {P{\'e}rez},
  {Benisty}, {Wilner}, {Baruteau}, {Bai}, \& {Ricci}}]{Zhang2018}
{Zhang}, S., {Zhu}, Z., {Huang}, J., {et~al.} 2018, ApJL, 869, L47,
  \dodoi{10.3847/2041-8213/aaf744}

\bibitem[{{Zhu} \& {Yang}(2021)}]{Zhu2021}
{Zhu}, Z., \& {Yang}, C.-C. 2021, \mnras, 501, 467,
  \dodoi{10.1093/mnras/staa3628}

\end{thebibliography}

%%%%%%%%%%%%%%%%% APPENDICES %%%%%%%%%%%%%%%%%%%%%
%\appendix

% \listofchanges
%%%%%%%%%%%%%%%%%%%%%%%%%%%%%%%%%%%%%%%%%%%%%%%%%%
\end{document}